# Computational Analysis of Climate Policy

by Carolyn Hicks

# Table of Contents

# Abstract

This thesis explores the impact of the Climate Emergency movement on local government climate policy, using computational methods. The Climate Emergency movement sought to accelerate climate action at local government level through the mechanism of Climate Emergency Declarations, resulting in a series of commitments from councils to treat climate change as an emergency.

With the aim of assessing the potential of current large language models to answer complex policy questions, I first built and configured a system named PALLM (Policy Analysis with a Large Language Model), using the OpenAI model GPT-4. This system is designed to apply a conceptual framework for climate emergency response plans to a dataset of climate policy documents. I validated the performance of this system with the help of local government policymakers, by generating analyses of the climate policies of 11 local governments in Victoria and assessing the policymakers' level of agreement with PALLM's responses. Having established that PALLM's performance is satisfactory, I used it to conduct a large-scale analysis of current policy documents from local governments in the state of Victoria, Australia. This thesis presents the methodology and results of this analysis, comparing the results for councils which have passed a Climate Emergency Declaration to those which did not.

This study finds that GPT-4 is capable of high-level policy analysis, with limitations including a lack of reliable attribution, and can also enable more nuanced analysis by researchers. Its use in this research shows that councils which have passed a Climate Emergency Declaration are more likely to have a recent and climate-specific policy, and show more attention to urgency, prioritisation, and equity and social justice, than councils which have not. It concludes that the ability to assess policy documents at scale opens up exciting new opportunities for policy researchers.

# Acknowledgements

I wish to thank my supervisors, Dr Kathryn Davidson and Dr Jey Han Lau, for bringing their wonderfully diverse knowledge and skills together to make this project possible. Both have been unfailingly positive, enthusiastic and supportive, and I couldn't have done it without them. I am especially grateful to Dr Thi Minh Phuong Nguyen for her support and her consistently thorough, specific and helpful advice.

I am very grateful to the Office for Environmental Programs and the University of Melbourne for enabling me to tackle such an exciting interdisciplinary project, and for approving all those enrolment variations! It has been a real joy to be able to use the knowledge I've acquired in such a cohesive and satisfying way.

Many thanks to the climate policymakers and implementers who volunteered their time for this project. Your insights were valuable and greatly appreciated, as is the important work you do every day.

I would like to acknowledge the late Philip Sutton for his dedication to the Climate Emergency movement, which played an important role in my interest in this space. I think he would have found the outcomes of this work both fascinating and frustrating.

Thanks to my colleagues at Evergen for providing the flexibility and headspace I needed to get this done. And to my family and friends for their love and support.

It is risky tackling a project such as this which combines two disciplines. My degree has involved subjects from multiple departments but has not been focused in any particular one, and this thesis is written in the hope of providing substantive value in the fields of both computer science and climate policy analysis. My supervisors have guided me on the conventions and requirements of their respective disciplines and I hope I have managed to meet the standards that both fields demand.

# Preface

## Declaration

The work in this project was undertaken in partial fulfilment of the requirements of the Master of Environment degree for the University of Melbourne.

This thesis comprises only my original work, and all other material has been duly acknowledged in the text. Any errors contained herein are solely my responsibility.

The views expressed are my own and do not necessarily reflect the views of the University of Melbourne or Office for Environmental Programs.

## Declaration of Generative AI usage

While this thesis contains a significant amount of text that has been generated by GPT-4, this has all been cited with "PALLM" as the author. I have occasionally interacted with ChatGPT to verify my understanding of a word or concept but have not used any of its output directly, nor have I asked it to review, structure or otherwise guide my writing. No other AI tools besides GPT models were used.

## Word count summary

| Preface, references, tables and appendix | 6937 |
| --- | --- |
| Literature review | 3285 |
| Introduction – Conclusion (excluding lit review) | 16962 |
| Complete document | 27184 |

# List of Figures

# List of Tables

# 1. Introduction

Local governments play an important role in responding to the climate crisis. As the closest level of government to local communities, they provide key services which can be severely affected by climate change, they have the potential to implement solutions on the ground to address climate change and its impacts, and they must be responsive to democratic pressure (Rosewarne, 2022b). The Climate Emergency movement makes strong demands for climate action at the local government level, and has had a highly visible impact around the world (Cedamia, 2024).

Trends in local government climate policy can indicate where action is being taken, where impacts are felt and where communities are demanding change (Fuhr et al, 2018). However, it is difficult to systematically survey the climate policy landscape, as policy actions and documents are not standardised. Assessing climate policy across multiple actors is a time- and labour-intensive task (Lamb et al, 2018; Davidson, 2020).

Interest has been growing among researchers in using automated tools to analyse policy documents through computational text analysis (Hsu & Rauber, 2021; Sachdeva et al, 2022; Davidson et al, 2024). New methods could map levels of ambition and innovative policy approaches, and allow tracking of changes over time. Various techniques have been used to perform computational analysis on policy documents, including topic modelling and logistic regression (Hsu & Rauber, 2021; Sachdeva et al, 2022), but to date, no significant published study has used the capabilities of large language models (LLMs) to analyse climate policy. In this study I aim to do so, using a conceptual framework developed for climate emergency response plans, and a technical solution based on GPT-4, a large language model produced by OpenAI.

The research questions of this study are:

- RQ1. What are the current capabilities of large language models in assessing policy documents?

- RQ2. To what extent has the language and priorities of the Climate Emergency movement influenced Australian local government policy?

I will address these questions through the following research objectives:

- RO1. Build and evaluate a software tool which uses a retrieval system based on GPT-4 to answer questions about climate policy documents (addressing RQ1)
- RO2. Evaluate the capabilities of large language models for assessing policy documents (addressing RQ1)
- RO3. Examine the extent to which the language and priorities of the Climate Emergency movement have influenced policy in Australian local governments (addressing RQ2).

The history of the Climate Emergency movement provides important context for this study. In 2016 the City of Darebin, in Victoria, Australia was the first government in the world to pass a Climate Emergency Declaration (CED) - a statement recognising that climate change is a global crisis requiring mobilisation of action and resources on an emergency scale, and committing the government to act with urgency to reduce emissions and address potential impacts. A small number of other local governments within Victoria and the USA followed suit in 2017 and early 2018, then in late 2018 a wave of CEDs began to spread around the world, fuelled by pressure from movements such as School Strike for Climate and Extinction Rebellion (Salvia et al, 2023; Soler-i-Martí et al, 2024). To date over 2200 governments have passed a CED, more than 1000 of them in 2019. The Climate Emergency movement specifically targeted local government (Spratt, 2019) in the hope that local governments and citizens acting in "emergency mode" (Salamon, 2019) would exert bottom-up pressure to enact greater change at a national and international level. Characteristics of "emergency mode" at the local government level include clear purpose and priority, attention to existential risks, strong leadership and fairness, together with "an integrated package of solutions for a safe-climate economy, zero emissions and large-scale carbon dioxide drawdown" (Spratt, 2019, p7).While the global wave of CEDs generated much discussion, its long-term impact is unclear. This study explores the impact of CEDs in Victoria on policy, by analysing a dataset of policy documents which includes policies from both CED councils (local governments which have passed a CED) and non-CED councils (which have not).

Following this introduction, Chapter Two of this study reviews the literature to date on local government climate policy, the Climate Emergency movement, the development of LLMs and the use of computational tools for policy analysis. It describes research which has analysed the content of CED documents and interviewed activists and policymakers, summarises how LLMs have been used to date in climate research, and provides an introduction to key concepts. Chapter Three describes the methodology for the four research objectives, including development and validation of the technical solution. Chapter Four details the results of the validation process, a large-scale analysis of the policy dataset, an investigation into variability of generated responses, and a qualitative assessment of policy alignment. Chapter Five discusses the strengths and limitations of the technical solution, and the findings of the analysis in the context of climate policy literature. Chapter Six concludes the thesis with recommendations for future research.

## 2. Literature Review

Every government at every level needs to address the problem of climate change, and each develops policy to suit its unique needs and circumstances. In Australia the climate policy landscape is highly fragmented, and the 566 local governments differ widely in their response to the issue. In recent years 115 local governments in Australia have made declarations of "climate emergency", along with more than 2200 governments worldwide (Cedamia, 2023). The climate emergency movement was fuelled by community desire for stronger climate action, but it's not clear what material impact it has had (Gudde et al, 2021; Greenfield et al, 2022; Davidson et al, 2024).

Shifts in the policy landscape are difficult to assess due to the variation in scope, approach and format of policies across local governments. Social science researchers are beginning to take advantage of recent developments in natural language processing (NLP) to conduct large-scale analysis of policy documents (see Hsu & Rauber, 2021; Sachdeva et al, 2022; Davidson et al, 2024).

This literature review will outline the current state of research with regard to the climate emergency movement's impact on local government climate policy, and methodologically relevant developments in NLP. Firstly, it will describe the current state of climate policy in Australian local governments and debates around the concept of climate emergency. Secondly, it will discuss relevant conceptual frameworks for policy analysis, and give a brief overview of research to date that uses computational tools for this purpose. Thirdly, it will review the development of large language models (LLMs) and introduce some concepts in the field of NLP which have relevance for policy analysis. Finally, it will describe some ways in which NLP researchers are using LLM capabilities in the climate space.

## 2.1 Climate policy in local government

Local governments play an important role in climate governance (Rosewarne, 2022a). They can actively pursue emissions reductions strategies in sectors such as housing and transport, and are often more responsive to community pressure due to their "direct and personal" relationship with their constituents (Rosewarne, 2022b, p227). A 2022 survey by the Climate Council of 158 Australian councils summarised their mitigation activities as relating to transport, renewable energy, the built environment, and home electrification (Cities Power Partnership, 2022). Current work by Davidson et al (2024) groups local government mitigation strategies into two categories: technological solutions such as LED lighting, electric vehicles and renewable energy; and a broader set of sustainable practices including agriculture, supply chains, access to nature and financial investment. This work also found many other patterns of discourse in council climate policies such as adaptation planning, social mobilisation, and linkages to other levels of government (Davidson et al, 2024).

Decarbonising the energy system offers particular opportunities for climate action at local government level. McGuirk et al (2014, p. 2723) noted in 2014 that a focus on energy efficiency as a climate issue, and the decentralisation of energy supply, were trends that could reposition local governments as "climate activists … innovators and experimenters". The ability to invest directly in renewable energy via power purchase agreements and small-scale generation has provided new opportunities for local governments to take direct and effective climate action, as has been studied in Australia (Rosewarne, 2022a), the UK (Kuzemko & Britton, 2020), and the US (Armstrong, 2021).

The "climate emergency" movement has focused attention on the role of local government in combating climate change. In 2016 the City of Darebin in Victoria, Australia was the first local government in the world to make a declaration of climate emergency (Davidson et al, 2021), and within four years more than 2000 governments worldwide had followed suit (Cedamia, 2023). Campaigners advocated for local governments to make climate emergency declarations in the belief that this could trigger a shift in governmental priorities, greater mobilisation of resources, and a departure from "business as usual" modes of operating (Salamon, 2019; Spratt, 2019).

The framing of climate change as an emergency has been critiqued from multiple perspectives, including the risk it poses to democratic processes, the potential to draw resources from other critical environmental and social issues, and a narrow and deterministic focus on deadlines (McHugh et al, 2021). A key focus of academic attention has been to assess the material consequences of climate emergency declarations through a close reading of declarations, climate action plans and related policy, and/or interviews with policy practitioners (Chou, 2020; Davidson et al, 2021; Howarth et al, 2021; Greenfield, 2022; Salvia et al, 2023). While these investigations are fruitful in terms of exploring the issues and tensions at play, no significant long-term effect from the act of declaring a climate emergency has yet been conclusively demonstrated (Howarth et al, 2021; Davidson et al, 2024).

## 2.2 Conceptual frameworks for climate policy analysis

The task of comparing climate policies across local governments is challenging due to the fact that there is no systematic way to make such comparisons. Each government produces (or doesn't) its own style of plan or policy, which may include specific targets and actions, or may consist mostly of rhetoric (Howarth et al, 2021). Several conceptual frameworks can assist in analysing local government climate action in the context of climate emergency. Zelli et al (2020) propose a theoretical framework for mapping the institutional complex of the climate-energy nexus, which addresses coherence and management (the way different institutions relate to each other and how this is formalised), and the impact of governmental interlinkages on the legitimacy and effectiveness of their governance. Kuzemko and Britton (2020) use 'capacity' as a lens for understanding local governments' approach to sustainable energy, enabling an assessment of a government's resources, relationships and motivations in this space. In relation to climate emergency declarations, Howarth et al (2021) consider four different pathways leading to an emergency declaration among London borough councils, and three categories of purpose: as a "statement of intent" to be followed with substantive action; as a symbolic gesture; and as a way to encourage local grassroots action (Howarth et al, 2021, p27). More specifically, Davidson et al (2020) propose a 'Climate Emergency Mode' (CEM) framework of 10 attributes, which include attention to specific climate adaptation and mitigation actions as well as governance criteria such as prioritisation of action, social mobilisation and equity.

The use of such frameworks requires a close and attentive reading of a large number of documents, entailing significant investment of time and resources. To assist and augment manual interpretation, computational tools for analysis of textual documents – techniques known as text mining, "text-as-data methods" (after Grimmer & Stewart, 2013) or computational text analysis methods (CTAM) – are increasingly widely used in the social sciences (Baden, 2022). Their application in political science and policy research is particularly useful given the importance of written communication in politics and the volume of data available; automated methods enable large-scale analysis of this data without requiring commensurately large-scale funding (Grimmer & Stewart, 2013). While topic modelling in particular has become a popular and accessible method for researchers, study designs and practices vary widely, and some authors have made recommendations on methodological practices to ensure robustness of findings, including comprehensive reporting of sources and methods, and a strong focus on validation (Müller-Hansen et al, 2020; Isoaho et al, 2021; Baden et al, 2022).

Several recent studies have used topic modelling and other computational tools to analyse climate policies and climate emergency declaration documents. Lamb et al (2018) used non-negative matrix factorisation to identify common topics in urban mitigation literature. Hsu & Rauber (2021) used topic modelling and network analysis to identify themes and relationships in climate actions in a large dataset of over 9000 actors, including cities, companies, regions and countries. They found a high degree of group similarity, and some evidence of orchestration of actions within subsets of actors, but also noted that opportunities to connect and share strategies had been missed. As well as topic modelling, Sachdeva et al (2022) used logistic regression to identify common factors among cities with net zero targets, and found four themes in language use which were linked with more ambitious targets: the use of specific metrics, identification of sources of emissions reduction, discussion of governance, and discussion of community engagement. Most recently, Davidson et al (2024) have used topic modelling to analyse climate policies from 196 Australian local governments, comparing those who did produce a climate emergency declaration (CED) with those who did not, and linking topics to the Climate Emergency Mode framework discussed earlier (Davidson et al, 2020). The topics generated in this analysis act as "policy framing patterns", finding that CED councils were more likely to propose policy focused on coordination and advocacy, as well as general sustainable practices, while non-CED councils had a stronger focus on technological solutions.

## 2.3 Large Language Models and Question Answering systems

While much of the use of computational tools for textual analysis in social science has focused on statistical methods such as unsupervised clustering, NLP has undergone rapid advances in recent years and a range of newer tools based on neural architectures are available (Zhao et al, 2023).

Practitioners of NLP aim to enable computational comprehension and generation of language. Zhao et al (2023) describe the development of the field, from statistical language models through deep learning with neural networks and the emergence of large language models (LLMs). The Transformer architecture, in which a self-attention mechanism enables the model to capture relationships in distant parts of input data (Vaswani et al, 2017), forms the basis of many modern NLP systems. Using this architecture, pre-trained language models such as BERT (Devlin et al, 2019) were able to perform well on a range of general NLP tasks without domain-specific training. As AI researchers (including the private organisation OpenAI) began increasing the scale of training data and computational resources to these models, large language models began to display what Zhao et al (2023, p2) call "surprising emergent abilities": for example, OpenAI's model GPT-3 was found to show high performance on 'few-shot learning', completing previously-unseen tasks when provided with natural language instructions and a small number of examples (Brown et al, 2020). After OpenAI's release of ChatGPT in 2022, which enabled non-technical users to interact with a GPT model using a conversational interface, there was a sharp increase in research papers relating to LLMs (Zhao et al, 2023).

Much NLP research has focused on the related tasks of reading comprehension and question answering (Rogers et al, 2021). "Reading comprehension" describes the ability of a system to read a given text and then summarise or otherwise refer to the content of the text in subsequent interactions; while "question answering" (QA) refers to a system which can accept natural language questions and provide appropriate responses. In combination, these capabilities describe a system which can process textual documents and then answer questions regarding their content. QA systems can be categorised as "open domain", where the model can answer questions on a range of topics or refer to various external sources, or "closed domain", where the model specialises in a specific domain of knowledge. Rogers et al (2021) provides a comprehensive taxonomy and overview of work in this area.

Daull et al (2023) describe the current state of LLMs with regard to "complex question-answering": a complex question is one with characteristics such as the need to access multiple sources of information, to decompose a question into multiple parts, or to use reasoning techniques such as deduction and induction. For example, answering the question "Who was president of the U.S. when superconductivity was discovered?" requires retrieving factual information in multiple steps guided by a logical strategy (Press et al, 2022, p5). An important aspect of complex QA is the ability to break down a complex question into sub-questions or steps, then integrate individual responses into a coherent answer. LLMs can sometimes demonstrate a "compositionality gap" (Press et al, 2022) in which the sub-questions are answered correctly yet the overall question is not. This gap can be addressed with techniques such as "Chain of Thought" (Wei et al, 2022) or "self-ask" prompting, demonstrated by example through few-shot prompting, which require the model to be more explicit and reflective about its answering process (Press et al, 2022). The Chain of Thought (CoT) technique breaks a task down into explicit steps before arriving at the final answer (Wei et al, 2022). In an example self-ask response to the above question, the model asks itself when superconductivity was discovered, and then incorporates the answer in a follow-up question asking who was president in that year (Press et al, 2022). Although CoT prompting increases the accuracy of final answers, it is not necessarily a genuine representation of the model's internal process, and may not reliably explain how the model arrived at its answer (Turpin et al, 2023).

Other limitations of LLMs include the tendency to "hallucinate", where generated text includes inaccuracies which either contradict the available internal information or cannot be verified by it (Zhao et al, 2023); lack of accuracy in questions with numeric and temporal responses (Tan et al, 2023); and lack of current knowledge due to outdated training data (Kraus et al, 2023).

To address some of these limitations, particularly for QA systems in domains where the accuracy and currency of available information are critical, a common strategy is to augment an LLM with access to an external knowledge source. An influential work on this by Lewis et al (2021) demonstrates a "retrieval-augmented generation" (RAG) model which combines the flexible generation ability of an LLM's parametric (internal) memory with the high degree of accuracy of a retrieval-based (external) information source. Zhu et al (2021) provide a thorough overview of the modern "retriever-reader" architecture for open-domain QA systems. In this architecture, a Retriever (IR or information retrieval system) queries an external data source and retrieves documents relevant to a given question. The documents are then processed by a Reader which either extracts answer spans from the documents, or generates a natural language response based on their content. The Reader system is generally a neural-based architecture specialising in reading comprehension.

## 2.4 LLMs in Climate Research

This is a fast-moving field of study with a great deal of work underway, much of it focusing on the capabilities of OpenAI's GPT models. Several researchers have identified the potential for retrieval-augmented generation systems to provide accurate, current information in scientific domains such as climate science, which could enable non-technical users to use natural language questions to access the latest scientific knowledge (Vaghefi et al, 2023; Kraus et al, 2023). Some examples of current work in this space include:

- "chatClimate" is a conversational AI chatbot based on GPT-4 with information sourced from IPCC reports (Vaghefi et al, 2023). The study investigated three scenarios: questions answered by GPT-4's internal knowledge, by reference to IPCC documents, and by a combination of the two. The hybrid approach was found to be most accurate by the climate experts who evaluated its responses.

- Kraus et al (2023) developed an LLM agent which utilises multiple external sources such as a Google search API and a dataset of ClimateWatch emissions data accessed via Python code. The agent uses a CoT strategy to set out the steps needed to answer a question, and can combine data sources to answer a complex multi-hop question. No evaluation was undertaken.

- CORE-GPT is a QA platform which answers questions in multiple scientific domains by generating a search query to retrieve research papers from an open-access database, then including the titles and abstracts in a GPT-4 prompt with instructions to answer the question solely based on their content (Pride et al, 2023). Evaluators ranked answers on comprehensiveness, trust and utility, finding that CORE-GPT scored over 8 out of 10 on these metrics in 75% of the question domains. This study also demonstrated that, when answering solely from internal knowledge, both GPT-3.5 and GPT-4 hallucinated over 70% of the citations they provided.

- "ChatReport" was designed by Ni et al (2023) to assess corporate sustainability reports using ChatGPT with reference to a reporting framework. While the study shows innovative methods including an "expert-involved prompt development loop" (p5), there does not appear to be robust evaluation of the system's output.
- Thulke et al (2024) have developed "ClimateGPT", a family of LLMs based on open-source models, which are trained to synthesise climate research using a RAG-based system with particular attention to interdisciplinary perspectives. The system was found to perform well using both automatic and human evaluation.

Rigorous evaluation of answers is critical in assessing the credibility of LLMs in reading comprehension and question answering. Automated evaluation of answers in QA systems has traditionally relied on 'lexical matching' against a predefined answer, which can be too strict and miss plausible answers that are expressed in different words (Kamalloo, 2023). Human judgement is considered the gold standard for evaluation (Kamalloo, 2023), but some researchers are experimenting with LLMs as evaluators. The G-Eval framework (Liu et al, 2023) uses GPT to evaluate texts that have been generated by an LLM for a task such as news article summarisation. First, a prompt is developed with instructions to complete an evaluation task by rating generated text on a particular metric such as coherence, according to specific criteria, on a numeric scale. GPT is used to augment this prompt with CoT steps for performing the evaluation, and produces a set of steps detailing how to assess text on this metric, such as "Check if the summary covers the main topic and key points of the news article, and if it presents them in a clear and logical order" (Liu et al, 2023, p3). For each metric to be evaluated, a scoring function calls GPT with four components of context: the initial prompt, the autogenerated CoT steps, the generated text to be evaluated (e.g. a summary), and the input context for the generated text (e.g. the original news article). Finally the scores are normalised. Liu et al's meta-evaluation of G-Eval showed that it outperforms other evaluation frameworks (by producing evaluations more consistent with human judgement), but notes that it appears to prefer GPT-generated summaries to human ones.

It is also very important for QA systems using external data sources to show the evidence supporting their answers. Bohnet et al (2022) propose a framework for measuring attribution and applying it to current LLMs. The authors note that attribution enables better human evaluation of system answers because it avoids the evaluator needing to judge the accuracy of a response directly. Recent work on automating systematic literature reviews with LLMs has developed new tools which enable auditing and verification of source information in LLM-generated text (Susnjak et al, 2024).

## 2.5 Conclusion

Climate policy at the local government level forms a critical part of the response to climate change, although the policy landscape is fragmented and difficult to analyse. Existing conceptual frameworks have the potential to enable systematic assessment of policies across governments, but this is difficult to achieve at scale. Computational tools could greatly assist in this endeavour.

This literature review has demonstrated that, to date, social science researchers have mainly made use of topic modelling and statistical methods to draw conclusions about the contents of policy documents, while NLP researchers have primarily used the power of LLMs to provide responses to factual questions relating to climate change. An opportunity exists to determine whether the natural language capabilities of LLMs can perform more sophisticated policy analyses than has been attempted using computational tools thus far.

An exploratory investigation of this opportunity could use an LLM-based retriever-reader pipeline to apply an existing conceptual framework to a dataset of policy documents. Careful attribution and evaluation would be necessary to ensure that document-specific answers are factual and well supported by evidence. If successful, the combination of LLM QA systems with policy-informed conceptual frameworks could enable nuanced large-scale policy analysis, providing a detailed and sophisticated view of the climate policy landscape.

# 3. Methodology

## 3.1 Introduction

The research objectives for this project are to develop and validate a technical solution for performing policy analysis assisted by GPT-4; to assess the capabilities of this solution; and to use it to examine the impact of the Climate Emergency movement on a dataset of climate policy documents. Underpinning the methodology for these objectives is the selection of a conceptual framework and development of questions that can identify the presence of framework attributes in a policy document.

Section 3.2 in this chapter describes the selected conceptual framework and the associated questions. Section 3.3 describes the development and final configuration of the technical solution. Section 3.4 describes the validation process and the methodology for analysis of its results, and section 3.5 describes the dataset of Victorian local government climate policy documents and related analytical methods.

## 3.2 Framework and questions

### Analytical framework

It was critical for this project to use a robust conceptual framework to underpin the technical solution and to ensure that the analysis generated by GPT-4 covers a comprehensive range of aspects of policy. A decision was made to focus on the Climate Emergency Mode (CEM) framework (Davidson et al, 2020), which was developed in the context of evaluating climate emergency response plans but can be usefully applied to any climate policy. The framework contains 10 attributes which examine various aspects of a policy document (see Table 1), including climate-specific and governance-related criteria. In order to use these attributes with a Question Answering system, the attributes needed to be expressed as evaluative questions, where a positive answer to the question indicates the presence of the attribute. The first set of questions used in initial development was taken from Davidson et al, 2024.

*Table 1: Climate Emergency Mode attributes and associated questions, from Davidson et al, 2024*

|    | Attribute | Description |
|----|-----------|-------------|
| 1  | Purpose of action | Does the plan truthfully state the 'what' and the 'why' of a climate emergency response? Does it plainly state who is responsible for action? Does it enable organisations as well as the communities they serve to mobilise behind a clear purpose of action? |
| 2  | Urgency of action | Does the plan necessitate rapid action? |
| 3  | Prioritisation of action | Have climate emergency actions have socio-governmental priority? Does the plan ensure that a climate emergency response is prioritised over policies incompatible with radical decarbonisation? |
| 4  | Institutional resource mobilisation | Does the plan allocate available discretionary funds and other institutional resources such as technical capacity and technological assets towards action to ensure delivery of the objective? |
| 5  | Social mobilisation | Does the plan actively empower the community to rally, support and work productively together to deliver climate action? |
| 6  | Restoring a safe climate | Does the plan include far-reaching mitigative efforts to restore a safe, decarbonised climate in conjunction with climate adaptation actions (occasionally within the resources referred to as 'building resilience' or 'reducing vulnerability') to rapidly address the causes of the climate emergency and lessen the impacts already being felt? Does it encourage societal, economic, environmental and cultural transformations such as sustainable practices and patterns of production and consumption such as new technologies and corresponding changes in markets, user practices, and policy responses? |
| 7  | Adapting to a changing climate | Does the plan include adaptation efforts to restore a safe climate to rapidly address the causes of the climate emergency and lessen the impacts already being felt? |
| 8  | Plan for informed action | Are the plan's targets, actions and monitoring based on current scientific data? Is there monitoring and evaluation capacity and research dedicated to closing knowledge gaps across varying aspects and developing critical solutions? |
| 9  | Coordination, partnerships and advocacy for action | Does the plan prescribe coordinated efforts between all sectors? Does it include advocating upward to state and national governments to support radical action? Building local capacity across council, their local communities and neighbouring local councils? |
| 10 | Equity and social justice | Given that the impacts of the climate emergency will have disproportionate impacts across society, does the plan ensure that both the burden of climate emergency action and the opportunities borne from a safe climate are equitably shared across local, national and even international communities? |

As can be seen in Table 1, the number of questions per attribute can vary. Some attributes require only one question to determine their presence, while others contain multiple aspects which must be explored with different questions.

The number and wording of the questions were iteratively revised, including during the validation process, in collaboration with two of the authors of Davidson et al, 2024. The reasons for these changes included:

- To simplify and clarify meaning
- To address aspects of the attribute which were not captured in the original set of questions, but which became apparent through document analysis
- To guide GPT-4 by referring specifically to climate action and asking for explicit references to the topic
- To incorporate feedback from policymakers during validation where questions were unclear, difficult to answer, or perceived to be irrelevant.

Table 2 presents three sets of questions which were used during validation. The initial questions in Table 1 were refined during development into question set A, which was used with the first evaluator. A small number of questions were added or removed to form question set B, which was used with seven evaluators. Finally, wording in some questions was adjusted further to create question set C, used with the final three evaluators. Each of the three question sets contained 20 questions in total. Question set C (highlighted) is the final version of the questions used in the large-scale analysis.

*Table 2: Questions associated with the Climate Emergency Mode framework attributes*

| N | Attribute | Question Set | Description |
|---|---|---|---|
| 1 | Purpose of action | A | Does the document explicitly define 'climate emergency' and if so, how? |

| | | A, B, C | Is climate action the core purpose or goal of the policy? |
|---|---|---|---|
| | | B, C | Does the document explicitly explain the need for action on climate change? |
| 2 | Urgency of action | A, B, C | Does the document explicitly call for rapid and urgent action on climate change? |
| | | A, B, C | Does the document give specific timeframes for its intended actions on climate? |
| 3 | Prioritisation of action | A, B, C | Does the document explicitly state that a climate emergency response must have higher priority than other policies? |
| | | A, B, C | Does the document explicitly state that all council activities must be aligned with climate policy? |
| 4 | Institutional resource mobilisation | A | Does the plan explicitly allocate funding (with a specific dollar amount) for climate action? |
| | | B, C | Does the plan explicitly allocate funding for climate action? |
| | | A, B, C | Does the plan explicitly allocate staff or other non-monetary institutional resources to climate action? |
| 5 | Social mobilisation | A, B, C | Does the document actively empower and educate the community to rally, support, and work productively together to deliver climate action? |
| 6 | Restoring a safe climate | A, B, C | Does the plan include specific actions for mitigation of greenhouse gas emissions, including technological solutions and behaviour change? |
| 7 | Adapting to a changing climate | A, B, C | Does the plan include specific actions for climate adaptation and resilience? |
| 8 | Plan for informed action | A | Are the document's climate targets, actions and monitoring based on current scientific data? |
| | | B | Are the document's climate targets, actions and monitoring based on current data? |
| | | C | Does the document provide well-sourced evidence to justify its climate targets and actions? |
| | | A | Does the plan aim to develop monitoring and evaluation capacities and research for its climate action? |

|   |   |   | B, C | Does the plan include specific measurable criteria to evaluate the success of its proposed actions? |
|---|---|---|------|---|
|   |   |   | B, C | Does the document describe plans to conduct research in the local community, to inform climate actions? |
|   |   |   | A, B, C | Does the document show evidence of innovation and policy experimentation in climate action? |
| 9 | Coordination, partnerships and advocacy for action | | A, B, C | Does the document show an explicit intent to advocate upward to state and national governments to support climate action? |
|   |   |   | A, B, C | Does the document explicitly encourage building local capacity across council, their local communities and neighbouring local councils for climate action? |
|   |   |   | A, B, C | Does the document refer to specific regional associations, alliances or other partnerships related to climate? |
| 10 | Equity and social justice | | A, B, C | Does the document explicitly discuss the impact of climate change on vulnerable communities? |
|   |   |   | A, B, C | Does the document explicitly discuss how to equitably share the benefits and opportunities of a safe climate? |

## 3.3 PALLM development

To fulfil the first research objective for this project, I have developed a technical solution for Question Answering with Retrieval Augmented Generation (QA/RAG), which can interact with GPT-4 to answer questions about the contents of policy documents. The system is named PALLM (Policy Analysis with a Large Language Model), which refers to the entire technical solution, including the questions and retrieval system, and the code which interacts with the GPT-4 application programming interface (API).

The development of PALLM was an iterative process in which multiple aspects were developed and fine-tuned simultaneously, including code design and structure, parameters of the retrieval system, the content of prompts and questions, the structure and format of the requested response, the scoring system, and the use of attribution.

## PALLM overview

PALLM is written in Python and uses the Langchain library to operate a QA/RAG system which reads and indexes policy documents, stores and retrieves relevant text, iterates over framework attributes and questions, interacts with GPT-4 and processes its responses.

PALLM is executed from the command line, taking PDF files as arguments plus additional parameters. It is designed to operate for a single council at a time, responding to all or some of the attributes in the CEM framework. When a single council has multiple relevant policy documents, PALLM reads and indexes all documents together, and then retrieves the text which is most relevant for each question across the set of documents. PALLM takes around three minutes to process a typical policy document and answer a set of 20 questions.

Figure 1 displays a high-level view of PALLM's operations, which are described in more detail in the following sections. To construct a complete analysis for a single council, first the relevant documents are processed and stored as embeddings in the retrieval system; then PALLM iterates over each question linked to the CEM framework attributes, sending a prompt to GPT-4 and receiving a generated response for each question. Finally, the responses are collated and an overall score for the council's policy is calculated.

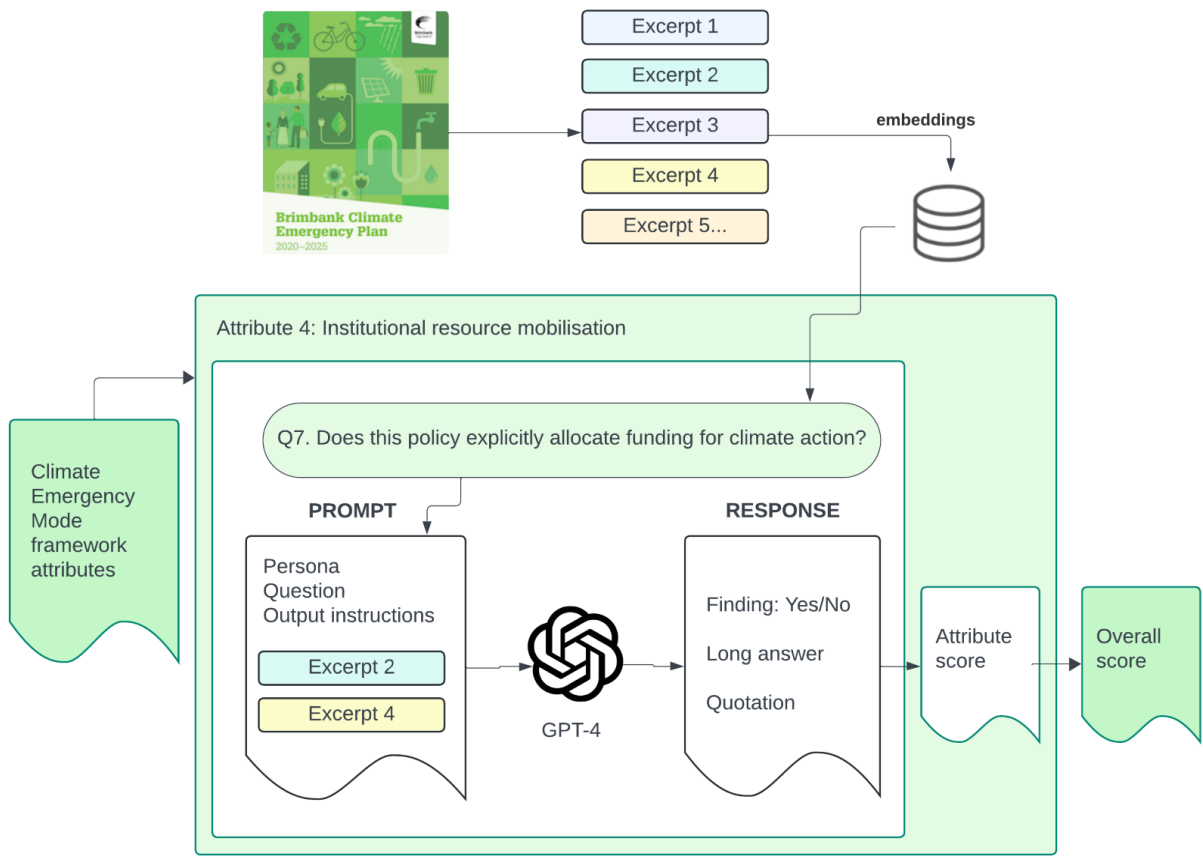

*Figure 1: The PALLM system, showing text extraction, retrieval and generation processes*

## Retrieval system

PALLM operates as a QA/RAG system, meaning that relevant excerpts are selected from policy documents for each question, and GPT-4 is instructed to answer the question with reference to the excerpts provided. The retrieval system is a critical component of PALLM because it determines which parts of the document GPT-4 has access to when answering questions.

For each document or set of documents to be analysed, the retrieval system is initialised as follows:

- Text is extracted directly from the PDF, using the PyMuPDFLoader library provided by the Langchain community. All formatting and images are removed.

- The text from all documents is split into chunks of 200 tokens (around 150 words) with an overlap of 10 tokens between each chunk.

- Each chunk is tokenised and converted to a series of embeddings. An embedding is a numerical representation of a token, which encapsulates the meanings and relationships associated with it. PALLM uses embeddings from the OpenAI model "text-embedding-ada-002".

- The chunks are written to a Chroma vector store, along with some metadata indicating their source document and page number.

During the analysis process, the following steps are repeated for each question:

- The full question text (including the attribute name) is converted to embeddings, and a query is made to the Chroma vector store to identify the stored document chunks with embeddings most similar to the question's.

- A maximum of 10 chunks is retrieved for each question.

- A prompt template is constructed which includes the original text of each chunk, plus instructions, question and format requirements.

### GPT-4 configuration

Development of PALLM took place over several months and initially used the GPT-3.5 model, but upgraded to GPT-4 once access became available. The validation and large-scale analysis in this study took place in March and April 2024 using the "gpt-4-0613" model. The large-scale analysis was conducted on April 13 and 14, 2024. This GPT-4 model has a context window of 8192 tokens, which means that the combined prompt and response for a single interaction must fit within roughly 6000 words (tokens are the individual units of meaning which an LLM processes). The model was not fine-tuned or customised in any way.

The only hyperparameter provided to GPT-4 was a sampling temperature of 0.0. The temperature parameter can range from 0 to 2 and lower values make GPT-4's output more predictable and consistent. An LLM constructs its output one token at a time by sampling from all words in its vocabulary and making a probabilistic selection. At higher temperatures, LLMs would make more 'creative' or 'random' choices. Lowering the temperature parameter increases the likelihood that the most probable next token will be selected, increasing the predictability of the generated output. It is important for PALLM to keep the temperature setting at its lowest possible value to ensure consistency of findings (see the discussion on Variability in section 4.4).

Although GPT-4 was used to provide PALLM's LLM capability in this project, it would be possible to use another OpenAI model or a chat-based LLM from a different technology provider in its place, as the API provided by Langchain can easily swap between models.

## Prompt structure

Each of the 20 questions in a question set (described in section 3.2) is presented to GPT-4 in a separate interaction. No memory or context is retained across interactions. The name of the attribute is prepended, so that the question incorporates that concept: for example, "With regard to urgency of action': Does the document give specific timeframes for its intended actions on climate?"

The prompt is in zero-shot format, as the use of examples during development seemed to distract the model from the source document. The prompt uses a 'persona' which positions the model as an analyst looking at local government policy and provides instructions on how to interpret and answer the question.

*Table 3: Prompt components*

| System prompt or 'persona' | You are a policy auditor performing a critical analysis of a local government policy. You are skeptical of vague generalities and require a high standard of evidence and specific detail to give a positive answer to a question. |
|---|---|
| Preamble | The policy states the local government's planned actions, and your analysis should focus on the aspects related to climate change. |

|  | Use the following pieces of context to answer a specific question. Only include information from the context in your response, and only answer the question in the affirmative when the context provides clear evidence. If the context doesn't answer the question, say so. |
|---|---|
| Context | [ up to 10 200-token excerpts from the document ] |
| Question | [ a question from the question set, with attribute name prepended ] |
| Format instructions | Give your answer in valid JSON format with the following keys:<br><br>   positive: [boolean] true if the question can be answered positively, false otherwise<br><br>   answer: [text] a critical analysis of the text, of about 250 words in length, responding to the question and giving supporting reasons<br><br>   quote: [text] if positive, include a brief quotation from the context which best illustrates the answer. This must be a direct quote from context. |

## Output format

Various types of response format were tried during development, including classification of answers into a three- or four-label typology. The results of classification were found to be unsatisfactory and complex to analyse, so the final version of the tool asks for a simple yes/no answer to the question. The prompt requests a response in JSON format, which allows PALLM to perform additional processing on it. The JSON structure contains three keys:

- "Positive": a boolean value indicating that the attribute in question is present in the document. I refer to this value as the 'finding', i.e. a "true finding" indicates that GPT-4 gave a positive answer to the question.

- "Answer": a paragraph of critical analysis on whether and how the attribute is present in the document. I refer to this value as the 'long answer'.

- "Quote": a quotation from the provided context which best demonstrates the attribute. This is only requested if the "positive" value is true, i.e. if the attribute is present.

## PALLM score

Once PALLM has completed analysis on all 20 questions in the set, it calculates a score out of 10. The scoring system counts positive answers to questions and normalises this across the number of questions per attribute. An attribute with a single question would receive 1 point if that question is answered in the affirmative; an attribute with four questions would receive 0.25 points for each true finding.

## Quotation verification

Attribution is a critical part of policy analysis, as it provides evidence for assertions about policy content. Several attempts were made to enable reliable attribution within PALLM, but this was not achieved to a satisfactory degree.

The prompt asks GPT-4 to include a quotation from one of the included chunks of context, which best provides evidence for the assertion that an attribute is present in the document. In most cases GPT-4 selects an appropriate excerpt from one item of context, but sometimes it concatenates sentences from different places, or if a partial sentence is selected from the beginning or end of a context item, GPT-4 may inaccurately extrapolate the rest of the sentence.

PALLM attempts to verify that the quotation provided in the model's response is valid, but within the scope of this study, no verification method was found to be sufficiently reliable.

An additional issue with attribution is that the long answer provided by PALLM may include phrases and sentences which are taken directly from the source document, but without acknowledgement. Hence while the quotation provided is often not faithful enough to the source text, the long answer may quote it verbatim without citation.

Given these technical limitations, the quotations provided by PALLM have not been included in results or discussion for this study. Any quotations from policies cited in this study have been taken directly from source documents.

## 3.4 Validation

Research objective RO1 requires validation of PALLM's performance. Validation took part in two stages, first internally during the process of development, and secondly through an external validation process with local government policymakers.

### Internal validation

During PALLM's development, I carried out a continual process of internal validation to evaluate the changes being made. This process involved identifying 'gold standard' answers for questions relating to a given policy, and comparing PALLM's responses to those. Initially these answers were sourced from published case studies (Davidson et al, 2020; Greenfield et al, 2022), which focused on policies that displayed many attributes of the framework; but it was also necessary to test PALLM's ability to detect lack of support for CEM attributes, and this required running it on policies with weaker commitment to climate action. To judge PALLM's performance on policies without a published case study, I first carefully read the policy, answered PALLM's questions manually, then generated an analysis and compared PALLM's responses to my manually written ones.

### External validation

The process of external validation required policy experts to assess the quality of PALLM's output, which necessitated a high level of familiarity with the policies being assessed. As it would be challenging to find participants skilled in policy analysis with the time available to closely analyse new documents, the selection process for PALLM evaluators targeted climate policymakers within Victorian local governments, who are already very familiar with the policies they work with.

In mid-February, 2024, I sent an email to the 41 local governments in Victoria who have issued a declaration of climate emergency (CED) (see Appendix), inviting them to participate in a research project involving AI-generated policy analysis. A total of 16 responses were received, and after initial discussion and signing of consent forms, 11 policymakers from 11 local governments (from regional and metropolitan areas) agreed to participate in an evaluation of PALLM. Between February 26 and March 20, six online interviews were conducted, and five participants chose to respond via an online survey form. The process was the same for both modes: I sought agreement from the evaluator on the primary climate document for the local government, then generated a PALLM analysis of this document, and the evaluator reviewed the analysis. Evaluators were asked to characterise their agreement with PALLM's answer to each of 20 questions about the policy, using a taxonomy of agree/disagree/unsure. For each question, evaluators had the opportunity to explain their choice or comment further. Evaluators were also asked for their general thoughts about PALLM's output, and if they believed such a tool could be useful to them in their future work.

Interviews typically took around 30-45 minutes, and often involved a highly engaged discussion. The online survey submissions did not provide as much detail in their responses.

When evaluation took place by interview, the generated analysis was sent by email at least one day prior to the interview. For the final three evaluators, I sent the questions a day earlier than the generated analysis, with the suggestion that evaluators should consider their own answers to the questions before reading the PALLM analysis. Even in those cases, there was no indication that evaluators explicitly considered how they personally would have answered the questions before reading the generated analysis.

Quotations from surveys and interviews in chapter 4 will be referenced with the codes given in Table 4, to preserve the anonymity of respondents.

*Table 4: Classification of councils which participated in validation, with reference codes*

| Council | Code |
| --- | --- |
| Regional North | I-S-1 |
| Metropolitan West | S-H-1 |

| Metropolitan North | I-M-1 |
| Regional North | S-M-1 |
| Metropolitan East | I-P-1 |
| Regional North | S-S-1 |
| Metropolitan East | S-M-2 |
| Metropolitan East | I-S-2 |
| Regional South | I-B-1 |
| Metropolitan South | I-M-2 |
| Regional City | S-B-1 |

## Fine-tuning of questions

As discussed in section 3.2, the questions used by PALLM were adjusted after the first interview and again after the eighth, resulting in three question sets. This made analysis of the validation results more complex, but addressed issues identified by evaluators.

## Analysis

The focus of analysis was a quantitative analysis of agreement rates, which was calculated in a spreadsheet. Of the 220 agreement choices made by evaluators (from 11 councils, across a set of 20 questions), 217 choices were available for analysis, as three had to be omitted due to question wording changes. The agreement rate was calculated by counting the number of PALLM findings which evaluators agreed with, expressed as a proportion of the total agreement choices counted for each evaluator. The changes in questions across evaluations mean that it is not possible to express a single agreement rate for the entire validation process, but the agreement rates per set are given in section 4.2. The agreement choices were also considered with regard to the respective PALLM finding, to calculate the proportion of disagree/unsure choices relative to positive or negative findings.

Finally, a qualitative assessment was undertaken of the evaluator's comments regarding PALLM's analysis, and of their responses to the two general questions, to identify common themes.

## Limitations

While the validation process was generally successful, with a good response rate and rich data for analysis, there are some limitations to its value, mainly related to the potential for bias on the evaluators' part.

Evaluators knew from the outset that the analysis was generated by AI, which can introduce bias (Kamalloo, 2023). A more robust process would have invited participation without specifying the source of the analysis, and could have included human-generated analyses along with PALLM-generated ones. This would have required considerably more resources to carry out, and may have reduced the response rate, as the involvement of AI generated particular interest among evaluators.

Evaluators did not necessarily consider their own answers to PALLM's questions, prior to reading the generated analysis. If they had been required to do so, they may have formed a different view to the one PALLM presented, rather than simply reading PALLM's response and finding it satisfactory.

Categorising agreement with "agree/disagree/unsure" labels is somewhat simplistic. Evaluators occasionally expressed the view that PALLM's finding was correct, but its long answer had missed important details, and the categorisation was unable to capture that.

Finally, evaluators who work directly with a policy may interpret a question as it relates to their organisation or broader context, rather than the specific contents of a document. An evaluator who has read the policy closely but has no other connection to it may take a more focused view, more appropriate for comparison with PALLM's. (This does not suggest that evaluators were biased in favour of their policy - see section 4.2.)

## 3.5 Analysis

The third research objective for this project involved building a dataset of recent climate policies from Victorian local governments and using PALLM to generate an analysis for each. This section describes the methodology for those actions, as well as an investigation into variability in PALLM, and other analytical methods used.

### Dataset

Earlier work by Davidson et al (2024) collected climate policy documents from all over Australia, which formed the basis for this project's dataset. Policies from outside Victoria were excluded from the dataset, and it was augmented as follows:

- A manual search was undertaken for a climate or sustainability plan for any Victorian council not already represented.

- While the previous work had excluded general sustainability strategies (due to their proportion of content unrelated to climate), this project has included them where no climate-specific document exists. The retrieval-based nature of PALLM means that only climate-relevant content is selected for analysis.

- When a council's most recent document dated from before 2020, a manual search was made to look for more recent policies.

This work was undertaken in March 2024, and does not include policies which were released after this time. A total of 96 documents were found, either climate- or sustainability-focused (where 'climate-focused' includes topics such as emissions reduction, adaptation and energy transition), across 73 councils. Multiple documents were included for a single council if they were released in or after 2020 and addressed different aspects of climate action. Documents released prior to 2020 were only included when no more recent document was available. The earliest document in the dataset is from 2010. One document was excluded due to being composed entirely of images and therefore unsuitable for text extraction, resulting in a total of 95 documents.

## Results

The large-scale analysis conducted on this dataset consisted of two independent PALLM analyses for each council, across the full dataset of documents. The output was recorded in JSON text files, comprising 2920 findings and associated long answers and quotations, from two sets of 20 questions across 73 councils. Findings that were inconsistent across the two executions were excluded from the results before scores were calculated. The data was then transformed into CSV format and analysed using Python with Pandas and Matplotlib libraries. The JSON output data is available in a Github repository at [https://github.com/xpatiate/nlp-climate-vic-output](https://github.com/xpatiate/nlp-climate-vic-output).

PALLM's scoring system is described in section 3.3. Scores are calculated for each of the 10 attributes in the CEM framework, and each council is assigned an overall score based on the sum of its attributes. These scores are presented in section 4.3, with data on findings for individual questions. PALLM scores and question data are categorised by the CED status of the council, i.e. whether the council which produced a policy had or had not passed a Climate Emergency Declaration.

### Variability

Careful attention was paid to the consistency of PALLM's responses, i.e. the variability of its findings across multiple analyses of the same document. With a temperature setting of 0.0, the expectation was that variability should be low, and initial experiments found this to be the case, but more variability than expected became apparent during large-scale analysis (see section 4.4 for further discussion). Repeated analyses for the 11 councils used in validation showed a variability rate of up to 11%. Experimentation with the 'top_p' hyperparameter did not improve this.

Observation of the long answers associated with differing findings showed that positive findings were more speculative, relying more on inference and assumptions, whereas negative findings focused more on overt statements in the document. This led to the following changes being made:

- A second sentence was added to the system prompt (see Table 3, section 3.3), instructing the model to require a high standard of evidence for a positive finding.
- A technical change was made to provide the system prompt as a separate 'system message' as described in the Langchain documentation. Previously the sentence specifying system behaviour had been included in the general prompt as a 'human message'.

With these two changes in place, the level of variability dropped to 1.5%, with 22 questions out of 1460 showing differing findings across two complete executions of the full dataset.

By directing GPT-4 to be more 'critical' in its analysis, these changes to the prompt increased the proportion of negative findings and lowered mean PALLM scores by around 5% relative to the prompt used for validation. This affects the relevance of the validation, because the technical solution which achieved high agreement rates in validation is different to the solution used for large-scale analysis. The change was considered to be worth making due to the importance of consistency across executions. Given that most disagree/unsure validation responses related to PALLM's positive findings, agreement rates could in fact have been improved by a higher proportion of negative findings.

## Other analysis methods

A small number of additional methods were used to investigate and summarise other aspects of the analysis presented in chapter 4.

*Level of confidence for questions*

Of the 20 questions in the question set used in the large-scale analysis, some appear more likely than others to elicit a clear and unambiguous answer from PALLM. The reasons for this will be discussed in chapter 5, but the validation and variability investigation provide some data to indicate which questions are more reliable. To surface this, a basic "level of confidence" measure was calculated, classifying each question as Tier 1, 2 or 3. The inputs to this measure are:

- The agreement rate for the question from validation, calculated as the number of times an evaluator agreed with PALLM's finding, divided by the number of times the exact question wording was used (between 3 and 11);
- Whether this question was answered inconsistently (with one positive and one negative finding) for any of the 73 councils in the two large-scale analysis executions.

The criteria for each grade are:

- **Tier 1**: Agreement rate greater than or equal to 90% AND no inconsistent findings
- **Tier 2**: Agreement rate less than 90% OR any inconsistent findings (not both)
- **Tier 3:** Agreement rate less than 90% AND any inconsistent findings

*Levenshtein ratio*

The Levenshtein ratio is used in section 4.4 to compare the similarity of long answers, and was calculated using the Python Levenshtein library. The Levenshtein distance algorithm calculates the number of changes required to transform one string into another: a higher value denotes more changes therefore lower similarity between the strings. The Levenshtein ratio expresses this distance relative to the length of the longest string, i.e. as a value between 0 and 1, where 0 represents identical strings and 1 represents complete dissimilarity.

*Log-likelihood ratio*

A log-likelihood ratio is used in section 4.4 to indicate whether certain words are more likely to be associated with true findings or false. This was calculated with the Python Keyness library, which uses an algorithm based on Rayson and Garside (2000). The ratio is a measure of the frequency of selected words in two corpora. If a given word appears an equal number of times in both corpora (relative to the size of each corpus), its log-likelihood ratio would be 0. A high log-likelihood ratio indicates that a word is more strongly represented in one corpus than another.

For the values in section 4.4, the calculation was derived as follows:

- all PALLM-generated long answers from the two large-scale analysis executions were combined into two corpora: one for long answers associated with true findings, another for those associated with false findings;
- the log-likelihood of all words in both corpora was calculated, as well as the absolute frequency of each word in each corpus;
- the frequency (proportionate to corpus size) was used to determine whether each word was over-represented in the 'true-finding' corpus or the 'false-finding' one;
- the log-likelihood and polarity for each word of interest was extracted from the dataset.

*Qualitative assessment*

Section 4.5 presents a qualitative assessment of the ways in which the attribute 'Prioritisation of action' is present in the climate policy dataset. This process used PALLM's long answers as a starting point, by exploring the answers associated with positive findings for the questions in this attribute. Having noted some common themes and examples in the generated answers, it was possible to quickly locate relevant source documents and passages. In this way, PALLM acts as a research enabler, by summarising and selecting relevant examples which assist the researcher in working with primary documents.

## 3.6 Summary

This chapter has described the methods used to construct a QA/RAG system interacting with GPT-4 for the purpose of climate policy analysis; as well as the validation of this system, and the methods with which it has been used to conduct a large-scale analysis of a set of documents. These methods were developed and refined iteratively, with internal validation guiding the technical solution's development. As described, some changes were made to PALLM during the validation process, and between validation and large-scale analysis. The following chapter presents the results of these processes.

# 4. Results

## 4.1 Introduction

This study uses PALLM, an LLM-based tool, to explore the influence of climate emergency declarations on climate policy in Victoria. To achieve this, I built, tested and refined PALLM, and conducted validation with a group of policymakers from Victorian councils (with methodology described in Chapter 3). These evaluators showed high rates of agreement with PALLM's output, described in section 4.2, and this forms the basis for its use in a large-scale analysis of climate policy documents. The results of the large-scale analysis are presented in section 4.3. Section 4.4 explores some patterns in PALLM's responses which are illustrated by its occasional inconsistent findings, and section 4.5 presents a qualitative assessment of the source documents regarding prioritisation of climate considerations.

It is important to note that, while evaluators showed high rates of agreement with the version of PALLM that was used during validation, the results of the large-scale analysis have not themselves been systematically evaluated and should not be considered precisely accurate. PALLM is designed for detection of broad patterns in the policy landscape, and while the scores assigned to councils by PALLM facilitate comparison, they are subject to limitations (examples of which are highlighted in this chapter) and are not presented as authoritative findings. Small differences between council or attribute scores are unlikely to be significant, and only results that show a large relative difference between cohorts are discussed in this chapter.

## 4.2 Validation

The validation process generated 217 responses categorising level of agreement with PALLM findings, from 11 evaluators in Victorian councils. Across the 217 responses (encompassing three question sets), evaluators agreed with GPT-4's answer 89.86% of the time (in 195 questions), disagreed 8.29% of the time (18 questions), and were unsure 1.84% of the time (4 questions).

In 30 of the 195 questions (15.38%) where the evaluator agreed with a true/false finding, they also noted some shortcomings in the long answer, for example that the answer missed some relevant content.

Of the 22 questions where evaluators disagreed or were unsure about PALLM's finding, 14 questions (63.64%) had been answered positively by PALLM, and 8 questions (36.36%) were answered negatively. As the questions are evaluative (meaning a "yes" answer is favourable), this indicates that when evaluators disagreed with PALLM's finding, their view could more often be characterised as feeling that PALLM's positive view of the policy was unwarranted, and less as PALLM failing to recognise a strength of the policy. While interesting, this aspect of the validation results is less relevant after post-validation adjustments (discussed in section 3.5 under 'Variability') increased PALLM's tendency to make negative findings.

The overall rates of agreement with PALLM's analysis are shown in Table 5.

*Table 5: Rate of evaluator agreement with PALLM analysis, over three different question sets*

| Question set | Number of evaluators | Agreement |
| --- | --- | --- |
| A | 1 | 78.95% |
| B | 7 | 90% |
| C | 3 | 93.33% |

The agreement rate does not reflect the score that PALLM's analysis gave to each policy, i.e. the rise in agreement between sets A and C does not indicate that PALLM was giving higher scores to the policies with question set C. The actual PALLM score calculated in these analyses is not reported here, because the relevant factor is whether the evaluator agreed with the answers, not the content of the answers themselves. The changes made to the question sets from A to B to C were intended to improve the quality of PALLM's responses, thus raise the rate of agreement, and it does appear that the changes had that effect.

The rates of agreement with PALLM's answers varied for different aspects of the analysis. Table 6 shows the rates of agreement for each attribute of the PALLM framework. Agreement rates were averaged across all questions linked to each attribute for the second column, and across only the final form of the questions (question set C) in the third column. Questions related to the high-level objectives and tone of the policy (attributes 1 and 2) had higher agreement rates than questions relating to more specific or technical aspects of policy documents, such as funding (attribute 4) or evidence base (attribute 8).

*Table 6: Rates of agreement averaged across questions for each framework attribute*

| Framework attribute | Agreement rate: all questions | Agreement rate: question set C |
| --- | --- | --- |
| Purpose of action | 95.45% | 95.45% |
| Urgency of action | 100% | 100% |
| Prioritisation of action | 95.45% | 95.45% |
| Institutional resource mobilisation | 81.21% | 88.89% |
| Social mobilisation | 100% | 100% |
| Restoring a safe climate | 100% | 100% |
| Adapting to a changing climate | 90.91% | 90.91% |
| Planning for informed action | 63.01% | 83.33% |
| Coordination, partnerships and advocacy | 93.94% | 93.94% |
| Equity and social justice | 86.36% | 86.36% |

## Themes

While the primary validation analysis was quantitative, based on agree/disagree/unsure responses, evaluators also had the opportunity to explain the reason for their answer or to offer other thoughts on PALLM's response. Qualitative analysis of these comments has identified some themes, which can be categorised as strengths, limitations and potential applications of PALLM.

*Strengths*

Evaluators found the primary strengths of PALLM's analysis to be its accurate summary of high-level information, and the selection of relevant examples.

### Accuracy

A common theme in interview and survey responses was an impression of accuracy in PALLM's responses. Comments reflecting this include:

- "This was a very accurate read, and exactly what we were trying to show in that document" (I-M-2)
- "That's spot on. I don't think I could really add to that answer to be honest" (I-B-1)
- "The answer describes the timeframes really well. Highlighting the commitment through timeframes and budget cycles, the answer also includes examples of actions which have specific timeframes" (S-H-1).

### Relevant examples

Some evaluators appreciated PALLM's ability to select examples highly relevant to the question topic from disparate parts of the document.

- "Again, this is something that is mentioned throughout, not specifically in one section. So it's done a really good job of bringing that together with very relevant examples." (I-M-2)
- "The answer clearly explains how the policy empowers and works with the community to deliver climate action. It provides a great example of an action that empowers the community." (S-H-1)

*Limitations*

Limitations noted by evaluators included selective use of examples, omitting relevant information, and a lack of specificity.

### Selective examples

Most evaluators noted that PALLM's answers did not provide a comprehensive list of relevant actions, instead citing only a few examples. The examples used were sometimes unexpected or tangential, e.g. when PALLM listed community food gardens as a mitigation action, the evaluator observed that "facilitating community food gardens is quite minor, it isn't going to be part of any tangible shift to emissions reductions" (I-M-1). One suggested that PALLM's response should always include the words "for example" (I-B-1), to avoid giving the impression that it was listing all relevant actions.

### Technical information omitted

PALLM's responses sometimes failed to mention relevant technical information, for example where targets or actions were supported by technical analysis such as climate modelling. The evaluator typically agreed with PALLM's finding, but would have expected the detailed evidence in the document to be mentioned.

- "The answer's totally accurate, but in terms of providing more detail and specific detail, then referencing like the Z-NET model and the marginal abatement curve would be the key things to have mentioned as well" (I-B-1).

On one occasion, the examples cited by PALLM were not relevant:

- "The long answer does not provide insight or relevant examples of adaptation, rather helping the community mitigate emissions. There are adaptation actions that could be referred to, but the answer does not mention them" (S-H-1).

### Missed information in images or tables

Because PALLM's analysis uses a text-based retrieval system, PALLM could not access data that was provided only in graphical form. Evaluators were aware of this and also noted that it sometimes failed to detect information that was formatted in tables.

- "It's not picking up any of the names we've got in that column in the table that's about potential collaborators and partners" (I-M-1).
- "I suspect maybe the reason why it hasn't picked up on that [modelling] is that's probably in images in the document as opposed to the text" (I-B-1).

### Misinterpretation of statements out of context

Evaluators noted that in some cases PALLM interpreted text as a direct policy commitment when it appeared in a different context, for example a quote from a community member, or a description of historical or future actions.

- "I think it's picked up on that [statement made in a] historical context and has taken that as the current context" (I-M-2).

In one instance, a policy describes developing real-time emissions monitoring as a long-term future possibility, and the PALLM analysis interprets this as evidence of commitment to data collection and monitoring. The evaluator noted that despite years of effort, real-time emissions monitoring was still not available, so should not be treated as an indicator of current practices (I-M-1).

### Not specific enough

Evaluators sometimes commented that PALLM gave general responses, lacking in detail: "This is a real AI-type answer because it's so vague" (I-B-1).

Responses to questions for the 'Planning for Informed Action' attribute were often criticised on this basis, especially questions about evaluation criteria.

- "I think its answer [that evaluation criteria are present] is correct. But ... while there are some evaluation criteria included in the action plan, certainly at a high level, I think more detail could have been provided, at least at an action level, to evaluate what success looks like for each of those actions, and when each action would be regarded as complete" (I-B-1).
- "That was one thing that came up in the internal audit on climate change, that our action plan doesn't have clear measurable evaluation criteria. So GPT thinks [we] do, and I would disagree" (I-S-1).

Relatedly, it was sometimes noted that PALLM's answers took a generous view of the policy, rather than being sufficiently critical. (Note that post-validation changes to the prompt, as described in section 3.5, addressed this limitation to some extent.)

- "It's a positive view, it's clearly not someone trying to pick holes in [the policy]" (I-M-1).
- "I think it's giving us a little bit more credit than... [is warranted]" (I-P-1).

*Potential applications*

Most evaluators were positive about the potential usefulness of a tool like PALLM in their work, both for internal communication and to enable comparison with other governments.

### Objectivity and consistency

Some evaluators took PALLM's positive assessment of their policy as impartial validation of their efforts. PALLM's interpretation of the policy was felt to be more objective than an internal assessment, and could be useful in communicating priorities to other parts of the organisation.

- "I'm really glad that it pulled [these examples] from the document the way it did, because, as I mentioned earlier, we really did try to have that social lens embedded throughout. And it's really picked those up" (I-M-2).
- "I think these policies can often be interpreted subjectively … so having this as a very non-biased lens was actually, I found, very, very useful" (I-M-2).

The consistency with which PALLM would conduct an analysis across multiple policies was seen to be a particular advantage.

- "For me, the ability to deploy a consistent methodology to do this type of research is probably as valuable as the detailed answer that comes with it." (I-P-1)

### Benchmarking

Several evaluators mentioned the usefulness of such a tool for benchmarking policies against those of other governments.

- "[It could help us understand] who else has got great policies? … What are they doing that I'm missing? Where's the gaps for us?" (I-M-1)
- "That comparison [to other governments] is important for us, especially as a global city in terms of benchmarking and in terms of making sure we're continually progressing." (I-M-2)

## 4.3 Large-scale analysis

### Dataset characteristics

The dataset described in section 3.5 consists of 95 policy documents, 74 of which specifically relate to climate policy, and the remaining 21 are general environmental strategies with climate-focused content. The documents represent 73 of the 79 local governments in Victoria, as six did not have any current climate or environmental policy apparent on their public website (see Appendix).

Of the 41 CED councils in Victoria, 39 (95.1%) have a climate policy which has been created or updated since their date of declaration. The mean year of publication for these councils' policies is 2020. A high proportion of these councils (37, or 90.24%) have at least one climate-specific policy document.

Of the 38 non-CED councils, the mean year of publication for the policies in the dataset is 2018. A lower proportion (20, or 52.63%) of these councils have climate-specific policies, six have no identifiable climate policy, and the remaining 12 include climate as part of a broader environmental strategy.

### Council results

The large-scale analysis found PALLM scores across 73 Victorian councils ranging from 30% to 97.5% with a mean of 70.0%. A higher score means that the council's relevant policies show more prevalence of the CEM framework attributes. As Figure 1 shows, CED councils are likely to have a higher score: among the 41 councils which had passed a climate emergency declaration, the mean PALLM score (as defined in section 3.3) was 76.0%, while in the remaining 32 councils, the mean was 62.4%.

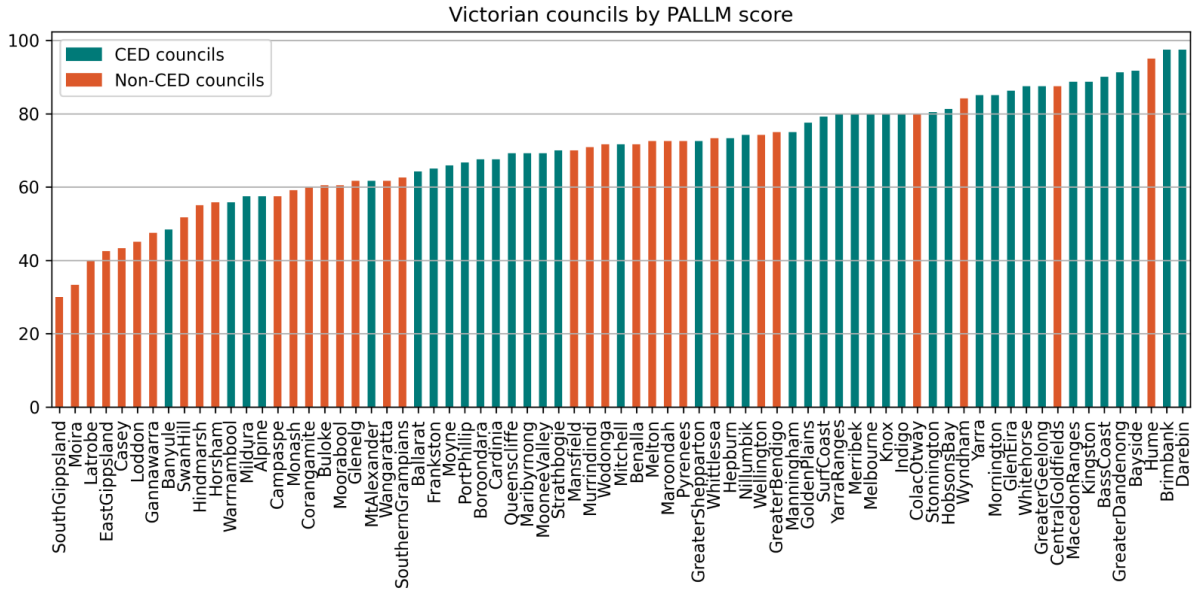

*Figure 2: PALLM score of Victorian councils, showing which have declared climate emergency*

## Attribute scores

When scores are grouped by each attribute of the PALLM framework, results show that CED councils score more highly on each attribute than non-CED (see Figure 2). Attributes where all councils score highly include 'Social mobilisation', 'Restoring a safe climate', and 'Adapting to a changing climate', while the lowest-scoring attributes are 'Prioritisation of action' and 'Equity and social justice'. Attributes where CED councils show a high relative difference over non-CED councils include 'Urgency of action', 'Prioritisation of action' and 'Equity and social justice'.

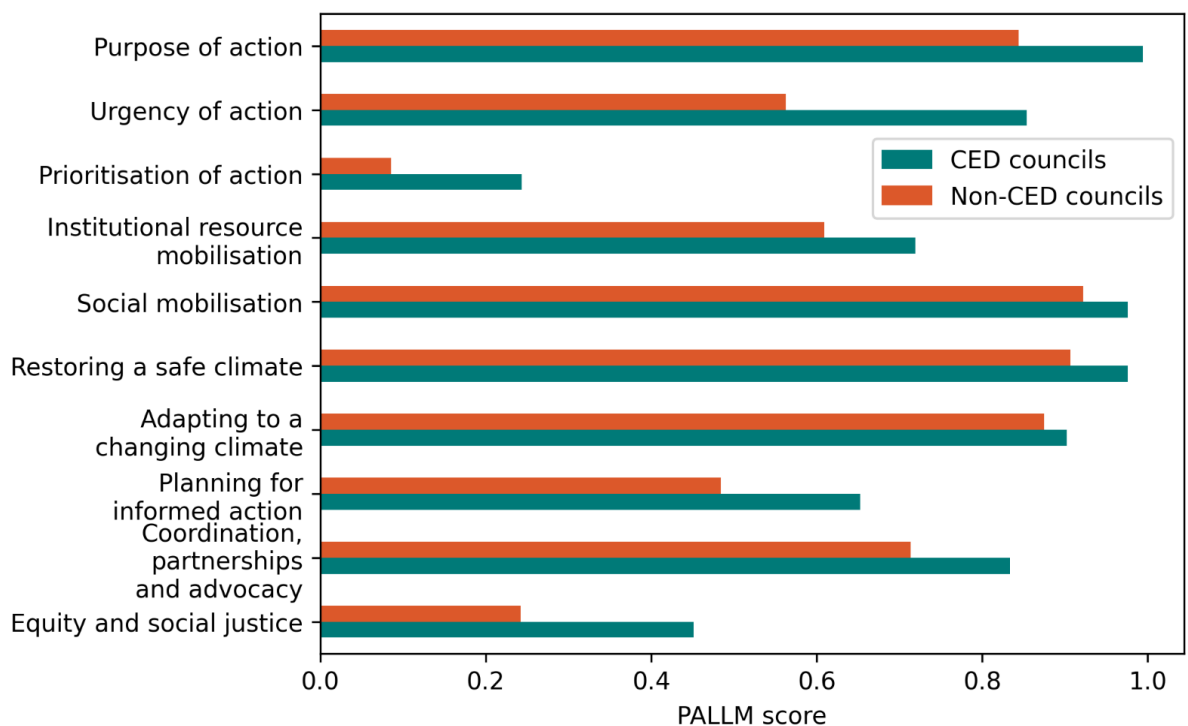

*Figure 3: Attributes from the PALLM framework, averaged across CED and non-CED councils*

## Attribute breakdown

The following section presents a detailed breakdown of the questions for each attribute, with examples from source documents and PALLM's long answers, and a level of confidence rating.

### 1. Purpose of action

*Table 7: Mean score for attribute 1*

| Mean score across all Councils | 93.7% |
| --- | --- |
| Mean for CED Councils | 100.0% |
| Mean for non-CED Councils | 85.5% |

*Table 8: Question details for attribute 1*

| Question | All Councils | CED Councils | Non-CED Councils | Level of confidence |
|---|---|---|---|---|
| 1. Is climate action the core purpose or goal of the policy? | 88.7% | 100.0% | 77.4% | Tier 2 |
| 2. Does the document explicitly explain the need for action on climate change? | 96.% | 100.0% | 93.5% | Tier 2 |

The 'Purpose of action' attribute relates to the high-level objective of a policy document. Because the dataset for this analysis consists of climate-related policy documents, it is unsurprising that most demonstrate climate action as their core purpose, as Table 7 shows. CEDs often include a commitment to produce a climate emergency response plan (Harvey-Scholes, 2019), and PALLM's answers for question 1 (see Table 8) indicate that 100% of CED councils have a policy document with climate as its core purpose. However, four of the CED councils in the dataset have documents with a more general title, such as Port Phillip's "Act and Adapt Sustainable Environment Strategy" (2024).

This discrepancy suggests that question 1 is not a good fit for a RAG system because it requires access to the full context of the document. PALLM's retrieval system provides GPT-4 with relevant excerpts from the document mentioning climate, but the question relates to the structure and predominant content of the document. Nevertheless, PALLM did accurately return negative findings for several councils which have a broader environmental strategy document. For example, in relation to East Gippsland Shire's "Environmental Sustainability Strategy" (East Gippsland Shire Council, 2022):

- "Climate action is a significant part of the policy, but it is not the sole focus. The policy aims to address a range of environmental and sustainability challenges prevalent across the municipality" (PALLM, East Gippsland, 2024-04-13a).

Most councils scored highly on question 2, demonstrating that most policies explain the context and need for climate action. For example:

- "[The document] highlights the adverse effects of climate change on the region's biodiversity, ecology, agriculture, tourism sectors, energy supply, urban form, and water supply. ... The document also aligns with the Paris Agreement commitment, which aims to keep global warming below 1.5°C higher than pre-industrial levels" (PALLM, Ballarat, 2024-04-13b).

*2. Urgency of action*

*Table 9: Mean score for attribute 2*

| Mean score across all Councils | 72.6% |
|---|---|
| Mean for CED Councils | 85.4% |
| Mean for non-CED Councils | 56.2% |

*Table 10: Question details for attribute 2*

| Question | All Councils | CED Councils | Non-CED Councils | Level of confidence |
|---|---|---|---|---|
| 3. Does the document explicitly call for rapid and urgent action on climate change? | 57.4% | 80.5% | 34.4% | Tier 1 |
| 4. Does the document give specific timeframes for its intended actions on climate? | 84.2% | 90.2% | 78.1% | Tier 1 |

Table 9 shows that CED councils are more than twice as likely as non-CED councils to explicitly call for rapid and urgent action on climate change. As noted above, most councils which passed a CED have since produced a document responding to that declaration of emergency, and this is clearly reflected in the language of urgency in the documents, for example: "We recognise that effective engagement and mobilisation of civil society and businesses in campaigning to demand emergency-mode action on climate change is critical" (Merri-bek City Council, 2022, p5). Non-CED councils are less likely to explicitly invoke a sense of urgency, as can be seen in Table 10.

Most councils regardless of declaration status were found to give specific timeframes for proposed actions. The timeline of intended action was sometimes mentioned by GPT-4 in responses to question 3, as support for a sense of urgency, for example:

- "The document indeed calls for rapid and urgent action on climate change. It outlines a series of priority actions, with high priority actions commencing within the next 12 months, medium priority actions by 2023, and low priority actions by 2025. This timeline indicates a sense of urgency in implementing these actions" (PALLM, Kingston, 2024-04-14a).

There is a high level of confidence in results for this attribute.

### 3. Prioritisation of action

*Table 11: Mean score for attribute 3*

| Mean score across all Councils | 17.2% |
|---|---|
| Mean for CED Councils | 24.4% |
| Mean for non-CED Councils | 7.9% |

*Table 12: Question details for attribute 3*

| Question | All Councils | CED Councils | Non-CED Councils | Level of confidence |
|---|---|---|---|---|
| 5. Does the document explicitly state that a climate emergency response must have higher priority than other policies? | 9.8% | 19.5% | 0.00% | Tier 1 |
| 6. Does the document explicitly state that all council activities must be aligned with climate policy? | 22.7% | 29.3% | 16.1% | Tier 2 |

'Prioritisation of action' was the attribute with the lowest prevalence across all councils at 24.4%, as shown in Table 11. PALLM's analysis found that only eight of the 73 councils in the dataset explicitly prioritise climate action over other policies, and as Table 12 shows, these eight were all CED councils. In these cases, PALLM's long answer sometimes referenced language from the original CED, or referred to structural factors regarding how the policy is to be integrated within the organisation:

- "The Brimbank Climate Emergency Plan 2020-2025 is described as sitting at the highest level of the Council's strategic planning framework, above and informing the four-year Council plans that are established at each new Council term. This indicates that the climate emergency response is prioritized above other policies." (PALLM, Brimbank, 2024-04-13a).

Question 6 explores whether climate policy is 'aligned' with other policies, which was found to be true in policies which clearly state that climate considerations should be addressed throughout council operations:

- "The documents provide clear evidence that all council activities must be aligned with climate policy. The council's plan outlines a comprehensive approach to addressing climate change, with every department within the council playing a role." (PALLM, Yarra Ranges, 2024-04-14a).

GPT-4 returned false findings on some policies which did state an aim to embed climate concerns throughout the organisation: for example,

- "These [new initiatives] include the establishment of a Climate Action Steering Committee, the implementation of a cultural change program, and the identification of best practice tools and techniques for embedding a culture of climate action within the council. The documents also mention the need for all staff to participate in these efforts and for climate action to become the new norm. However, these are general statements about the council's intentions and do not constitute a clear directive that all council activities must be aligned with climate policy" (PALLM, Banyule, 2024-04-13a).

This contributes to a lower level of confidence for this question. See section 4.5 for further analysis of this attribute based on qualitative assessment of the source documents.

*4. Institutional resource mobilisation*

*Table 13: Mean score for attribute 4*

| Mean score across all Councils | 67.1% |
|---|---|
| Mean for CED Councils | 72.0% |
| Mean for non-CED Councils | 60.9% |

*Table 14: Question details for attribute 4*

| Question | All Councils | CED Councils | Non-CED Councils | Level of confidence |
|---|---|---|---|---|
| 7. Does the plan explicitly allocate funding for climate action? | 55.7% | 70.7% | 40.6% | Tier 2 |
| 8. Does the plan explicitly allocate staff or other non-monetary institutional resources to climate action? | 77.2% | 73.2% | 81.2% | Tier 2 |

Table 13 shows that the allocation of monetary funding was found to be more prevalent in CED councils than non-CED. Many policy documents present proposed actions with a code indicating a funding category: for example, Colac-Otway Shire's "Climate Change Action Plan" (2023) specifies a cost category of Low, Medium or High for each action, as well as an indicator of resourcing given as 'F' (existing funding), 'S' (external funding) or both. PALLM was effective at interpreting and summarising this information:

- "The plan ... uses broad categories to indicate the cost of each action, such as 'Low ($0–$50,000)', 'Med ($50,000–$150,000)', and 'High (150,000+)'. The plan also indicates whether each action is funded within existing resources or subject to external funding and/or funding by Council as part of an annual budget process." (PALLM, Colac-Otway, 2024-04-13b)

The proportion of true findings was much closer for CED and non-CED councils in the allocation of non-monetary resources – 77.2% of all policies were found to allocate staff time and expertise to climate action. Overall, Table 14 shows that two-thirds of all councils were found to be allocating resources to climate action.

*5. Social mobilisation*

*Table 15: Mean score for attribute 5*

| Mean score across all Councils | 95.6% |
|---|---|
| Mean for CED Councils | 97.6% |
| Mean for non-CED Councils | 93.5% |

*Table 16: Question details for attribute 5*

| Question | All Councils | CED Councils | Non-CED Councils | Level of confidence |
|---|---|---|---|---|
| 9. Does the document actively empower and educate the community to rally, support, and work productively together to deliver climate action? | 95.6% | 97.6% | 93.5% | Tier 2 |

Almost all policies for all councils were found to discuss community engagement and education, as tables 15 and 16 show, though the level of detail varies greatly. For example, the City of Greater Bendigo's "Climate Change and Environment Strategy" (2021) includes a flagship project called "The Greater Bendigo Climate Collaboration", which aims to work with 1000 local households and 100 local businesses to make zero carbon plans, as well as supporting community action in many other ways. In contrast, while the Loddon Shire's "Environmental Sustainability Strategy Action Plan" (2013) does refer to community education (and PALLM considers this attribute to be present in the document), this is limited to the provision of informative material about sustainability on the Council's website.

PALLM's responses listed many proposed projects and actions related to community participation:

- "The policy outlines a range of initiatives aimed at fostering community engagement and action on climate change. These include the development and implementation of a community climate education plan and education programs for residents, businesses, and industry on a range of climate topics (CE13). The plan also includes the development of a community climate civic participation and leadership program, focused on increasing skills and ability for the community to act on climate (CE16). Furthermore, the policy encourages and supports community groups that use Council facilities to develop and implement a climate emergency plan (CE18)." (PALLM, Maribyrnong, 2024-04-13b).

### 6. Restoring a safe climate

*Table 17: Mean score for attribute 6*

| Mean score across all Councils | 94.1% |
| --- | --- |
| Mean for CED Councils | 97.6% |
| Mean for non-CED Councils | 90.6% |

*Table 18: Question details for attribute 6*

| Question | All Councils | CED Councils | Non-CED Councils | Level of confidence |
| --- | --- | --- | --- | --- |
| 10. Does the plan include specific actions for mitigation of greenhouse gas emissions, including technological solutions and behaviour change? | 94.1% | 97.6% | 90.6% | Tier 1 |

Tables 17 and 18 show that almost all councils were found to include mitigation actions in their climate policy documents, as would be expected. Of the four councils with a negative finding, three had only adaptation-specific policy documents present in the dataset. Typical mitigation actions included support for zero-emissions vehicles, generation and purchase of renewable energy, and improving energy efficiency in local buildings.

PALLM's responses listed many examples of mitigation actions and often categorised them using the suggested taxonomy of 'technological solutions' and 'behaviour change':

- "The plan includes technological solutions such as transitioning to all-electric and zero carbon energy, building and retrofitting homes and infrastructure to make them sustainable and climate resilient, and switching to more sustainable transport like walking, cycling, ride sharing and electric vehicles. The plan also encourages behaviour change, such as buying less, recycling and reusing more to achieve zero waste, and supporting sustainable and local businesses. The plan also mentions the integration of climate risks into corporate processes, which can be seen as a form of institutional behaviour change" (PALLM, Bass Coast, 2024-04-13a).

GPT-4 sometimes added explanatory comments which did not appear in the original document, such as noting that home composting "can reduce the amount of organic waste sent to landfill and thus decrease methane emissions" (PALLM, Central Goldfields, 2024-04-13a).

It also restated the categorisation of technological solutions/behaviour change in a novel way:

- "The plan's approach to mitigation is comprehensive, addressing both the supply (through technological solutions) and demand (through behaviour change) sides of greenhouse gas emissions" (PALLM, Central Goldfields, 2024-04-13a).

*7. Adapting to a changing climate*

*Table 19: Mean score for attribute 7*

| Mean score across all Councils | 88.9% |
|---|---|
| Mean for CED Councils | 90.2% |
| Mean for non-CED Councils | 87.5% |

*Table 20: Question details for attribute 7*

| Question | All Councils | CED Councils | Non-CED Councils | Level of confidence |
|---|---|---|---|---|
| 11. Does the plan include specific actions for climate adaptation and resilience? | 88.9% | 90.2% | 87.5% | Tier 1 |

Most councils were found to include adaptation actions in their climate policy documents, with little difference between CED and non-CED in the prevalence of this attribute (as shown in Tables 19 and 20). As an example, Whittlesea City's "Climate Change Plan" (2022) includes a wide range of adaptation actions including flood mapping, discouraging new settlements in hazard-prone areas, and tree planting to reduce heat risk. For councils with negative findings, adaptation may be discussed in forthcoming policy documents which do not appear in the dataset.

While mitigation and adaptation measures can of course overlap, GPT-4 seems generally able to distinguish between them, and did not consider sustainability actions that were not framed as adaptation as counting towards this attribute:

- "While some actions, such as advocating for large-scale renewable energy projects and improving water security, could indirectly contribute to climate resilience, they are not framed in terms of climate adaptation" (PALLM, Loddon, 2024-04-13b).

In policies where no adaptation actions were found, GPT-4 sometimes listed hypothetical examples, demonstrating its parametric knowledge of climate adaptation:

- "Adaptation strategies typically involve measures to deal with the effects of climate change, such as infrastructure improvements to handle increased flooding or heatwaves, or programs to protect ecosystems and biodiversity. Resilience strategies often involve strengthening community and ecological systems to withstand and recover from climate impacts" (PALLM, Merri-bek, 2024-04-13a).

## 8. Planning for informed action

*Table 21: Mean score for attribute 8*

| Mean score across all Councils | 58.2% |
|---|---|
| Mean for CED Councils | 66.0% |
| Mean for non-CED Councils | 48.4% |

*Table 22: Question details for attribute 8*

| Question | All Councils | CED Councils | Non-CED Councils | Level of confidence |
|---|---|---|---|---|
| 12. Does the document provide well-sourced evidence to justify its climate targets and actions? | 27.2% | 24.3% | 30.0% | Tier 3 |
| 13. Does the plan include specific measurable criteria to evaluate the success of its proposed actions? | 77.8% | 90.0% | 65.6% | Tier 3 |
| 14. Does the document describe plans to conduct research in the local community, to inform climate actions? | 67.2% | 79.5% | 54.8% | Tier 2 |
| 15. Does the document show evidence of innovation and policy experimentation in climate action? | 54.7% | 67.5% | 41.9% | Tier 2 |

The 'Planning for informed action' attribute showed relatively low prevalence across councils, as seen in Table 21. The questions in this attribute have a low level of confidence (see Table 22) owing to more inconsistent findings for questions in this group. Answers to these questions depend on a degree of judgement regarding, for example, what counts as 'innovation and policy experimentation' or 'well-sourced evidence'.

PALLM responses to questions in this category sometimes seem to base a true finding on stated intentions, rather than explicit evidence. For example, PALLM found that the Ballarat City Council's "Net Zero Emissions Plan" (2022) does include specific measurable criteria for evaluation because "The plan outlines a monitoring and evaluation framework that will be developed to assess the uptake and effectiveness of specific actions" (PALLM, Ballarat, 2024-04-13a). The document states that such a framework will be developed in future, but does not 'outline' it in any detail, and does not include evaluation criteria in the plan itself.

Even so, it is apparent that this attribute is prominent in some policies where proposed actions are backed by evidence and engage with forward-looking emissions reduction techniques. The Bass Coast Shire's "Climate Change Action Plan" (2021) discusses innovative projects such as teal and blue carbon sequestration, and provides extensive information on the modelling that underlies its targets.

## 9. Coordination, partnerships and advocacy

*Table 23: Mean score for attribute 9*

| Mean score across all Councils | 78.6% |
| --- | --- |
| Mean for CED Councils | 84.2% |
| Mean for non-CED Councils | 71.6% |

*Table 24: Question details for attribute 9*

| Question | All Councils | CED Councils | Non-CED Councils | Level of confidence |
| --- | --- | --- | --- | --- |
| 16. Does the document show an explicit intent to advocate upward to state and national governments to support climate action? | 67.4% | 80.00% | 54.8% | Tier 2 |

| 17. Does the document explicitly encourage building local capacity across council, their local communities and neighbouring local councils for climate action? | 89.% | 97.% | 81.2% | Tier 2 |
| --- | --- | --- | --- | --- |
| 18. Does the document refer to specific regional associations, alliances or other partnerships related to climate? | 76.9% | 75.6% | 78.1% | Tier 2 |

All councils showed relatively high prevalence of the 'Coordination, partnerships and advocacy' attribute (see Table 23). Advocacy to higher levels of government for stronger climate action was a key demand of the Climate Emergency movement (Spratt, 2019), and PALLM has found that CED councils are more likely to state such an intention than non-CED, as shown in Table 24. However, the level of confidence in this finding is only Tier 2, due to inconsistent findings across analyses. One PALLM response detects implied advocacy:

- "The policy outlines a commitment to work in partnership with other agencies and landowners to reduce fire risk to communities, which implies a level of advocacy to higher levels of government" (PALLM, Hepburn, 2024-04-13a).

Another counts advocacy to other organisations as advocacy to government:

- "While the document does not explicitly mention advocating upward to state and national governments, the inclusion of 'agencies' and 'research institutes' in their advocacy efforts could be interpreted as such" (PALLM, Corangamite, 2024-04-13a).

However, most true findings for this question are supported by a detailed answer which lists specific references to advocacy in the policy, and this occurs at a higher level for CED councils than non-CED. The Brimbank City Council "Climate Emergency Plan" lists specific goals and targets for advocacy at the state level, including a 100% renewable energy target and a price on carbon pollution (2020, p15).

For the remaining two questions in this attribute, both CED and non-CED councils score highly, indicating that most climate policies recognise the importance of partnerships and regional alliances. Every local government in Victoria belongs to one of eight Greenhouse Alliance organisations, which are frequently mentioned in the documents. The "Zero Carbon Merri-bek" plan (Merri-bek City Council, 2022) lists possible collaborators for every action, covering a wide range of organisations and sectors.

### 10. Equity and social justice

*Table 25: Mean score for attribute 10*

| Mean score across all Councils | 35.9% |
|---|---|
| Mean for CED Councils | 45.1% |
| Mean for non-CED Councils | 23.8% |

*Table 26: Question details for attribute 10*

| Question | All Councils | CED Councils | Non-CED Councils | Level of confidence |
|---|---|---|---|---|
| 19. Does the document explicitly discuss the impact of climate change on vulnerable communities? | 45.2% | 56.1% | 34.4% | Tier 1 |
| 20. Does the document explicitly discuss how to equitably share the benefits and opportunities of a safe climate? | 23.5% | 34.1% | 12.9% | Tier 3 |

Table 25 shows that discussion of equity and social justice is infrequent across councils, but relatively more prevalent in the CED cohort. PALLM found that the inequitable impact of climate change on vulnerable communities was discussed in 56.1% of CED council policies – 46% more often than in non-CED, as seen in Table 26. There is a high level of confidence in this result, with a high level of agreement from validation and no inconsistent findings. As an example, the Yarra City "Climate Emergency Plan" contains a specific action to "support vulnerable communities" (2020, p37), by means such as targeted communications, building capacity in community organisations, and assisting with home upgrades.

PALLM's answers typically list the characteristics of vulnerable communities mentioned in the policy, and cite proposed actions to support those communities:

- "The plan identifies those at greatest risk, including older people, infants/young children, those with existing medical conditions, and people taking medications that may affect their reaction to heat. The plan also includes measures to protect these vulnerable groups, such as flood mitigation measures, investment in resilient infrastructure, and the implementation of a heatwave plan." (PALLM, Pyrenees, 2024-04-13b)

The second question in this attribute has a low level of confidence due to some ambiguity in its interpretation, and was found to be true for only 23.5% of councils. In some cases PALLM's responses were very similar to question 19, focusing on negative impacts on vulnerable communities, but in others it highlighted actions which relate to equitable distribution of benefits of climate action:

- "This includes offering subsidised home energy assessments and supporting access to affordable energy efficiency upgrades. These actions indicate a clear intention to ensure that the benefits and opportunities of a safe climate are shared equitably across the community, including among those who are most vulnerable" (PALLM, Bayside, 2024-04-13a).

## 4.4 Variability

The PALLM large-scale analysis was executed twice on each of the 73 councils in the dataset, and the results were compared to establish consistency. Because GPT-4 is a privately maintained system, the way it functions could alter at any time, and its output hypothetically could vary due to changes in its internal operation. Excluding this theoretical possibility, given two executions of the same questions on the same input, with the 'temperature' setting at 0.0 (as discussed in section 3.3), it is reasonable to expect a high level of consistency between outputs. Complete consistency is not expected because the nature of parallel computation in graphical processing units introduces a degree of non-determinism in LLM behaviour (Kim et al, 2023).

The document excerpts provided to GPT-4 by the retrieval system will be the same on each execution, because these are deterministically selected based on the calculated similarity of the question to the stored document excerpts. Although not systematically verified in this study, manual inspection indicates that the same excerpts are selected and passed to GPT-4 every time a given question is asked in relation to a document. Therefore, any variability in the response is introduced within GPT-4, not in the input that is provided to it.

While variability across executions reached 11% in early iterations of PALLM, the final iteration of the tool had reduced this to 1.5% (due to improvements discussed in section 3.5). Across two large-scale analysis executions, each consisting of 20 questions for each of 73 councils with a total of 1460 questions, 22 of these questions had different true/false findings. Repeated executions on a subset of the dataset showed similar or lower levels of variability.

The figure of 1.5% variability refers only to the true/false finding in a PALLM response. Long answers can of course vary in more subtle ways, and usually do – only 4.7% of paired long answers with the same finding (two long answers to the same question for the same set of documents, generated on separate executions) are identical. Variability between long answers is better calculated with a measure of text similarity such as the Levenshtein distance, described in section 3.5 (in which a lower value denotes less similarity between two passages). The mean Levenshtein ratio for paired long answers with the same finding is 0.77, and varies from 0.49 to 1.0.

Although paired long answers can show significant lexical differences, these differences generally do not substantially alter the meaning of the passage. Table 27 shows two samples of paired long answers. The first sample illustrates the minimum ratio (i.e. highest level of variation) between a pair of answers, and the second sample illustrates the median level of variation. Different examples may be selected, common words or phrases may appear in different order, or similar ideas may be expressed in slightly different language.

*Table 27: Paired long answers with the same finding (PALLM, Campaspe, 2024-04-13; PALLM, Yarra, 2024-04-14)*

| Levenshtein ratio | Finding: true | Finding: true |
|---|---|---|
| 0.49 (minimum) | The documents provide evidence that the local government is actively empowering and educating the community to rally, support, and work productively together to deliver climate action. The policy outlines a multi-level collaboration strategy that includes households, organisations, city-wide projects, and regional advocacy. Households are encouraged to create climate action plans and participate in local community climate projects. Organisations, including businesses and schools, are supported to create climate action plans and commit to a 2030 target. City-wide projects aim to bring the community together to fund collective climate projects. Regional advocacy involves councils from the Loddon Campaspe regional network to create and design regional-scale climate projects. The policy also encourages the community to participate in the Greater Bendigo Climate Collaboration, retrofit homes with energy-efficient appliances, support local businesses to transition to renewable electricity, and support locally owned renewable energy generation and storage projects. | The documents provide evidence of a comprehensive approach to social mobilisation for climate action. The local government is actively engaging households, businesses, and organisations in the Greater Bendigo Climate Collaboration, encouraging them to develop and implement zero carbon plans. The policy also outlines support for local businesses to transition to 100% renewable electricity and increased energy efficiency. The community is encouraged to participate in local climate projects, retrofit homes with energy-efficient appliances, and support locally owned renewable energy generation and storage projects. The policy also advocates for sustainable transport and food systems, and it outlines plans for climate risk and vulnerability assessments. These actions collectively empower and educate the community to rally, support, and work productively together to deliver climate action. |
| 0.76 (median) | The policy does not explicitly state that climate action is the core purpose or goal. While the policy does address climate change, it | The policy documents do not explicitly state that climate action is the core purpose or goal of the policy. While the policy does address climate change |

| | does so as part of a broader environmental sustainability strategy. The policy's stated purpose is to guide the local government's decision-making and management practices to achieve environmentally responsible and sustainable outcomes for the municipality. It also aims to address the potential adverse impacts of climate change and promote the adoption of environmentally sustainable practices within the municipality. However, these are components of the policy's broader environmental sustainability focus, rather than being the core purpose or goal. | and outlines actions related to it, it is not clear that this is the central focus. The policy's stated purpose is to guide the local government's decision-making and management practices to achieve environmentally responsible and sustainable outcomes for the municipality. It also aims to address the potential adverse impacts of climate change and promote the adoption of environmentally sustainable practices within the municipality. However, these goals are part of a broader environmental sustainability strategy, which also includes other aspects such as reducing environmental degradation, building resilience, and promoting sustainable water use. Therefore, while climate action is a significant part of the policy, it cannot be said to be the core purpose or goal based on the provided documents. |

When PALLM gives a different true/false finding to the same question in relation to the same document, its long answers unsurprisingly have a lower degree of similarity than when the finding is the same. The mean similarity among paired long answers with different findings is 0.54, ranging from 0.47 to 0.63. The samples in Table 28 provide an interesting illustration of GPT-4 making two opposing arguments based on the same input.

*Table 28: Paired long answers with opposite findings (PALLM, Southern Grampians, 2024-04-14; PALLM, Macedon Ranges, 2024-04-13)*

| Levenshtein ratio | Finding: false | Finding: true |
| --- | --- | --- |
| 0.48 (minimum) | The documents do not provide well-sourced evidence to justify its climate targets and actions. While the documents outline a range of actions and strategies to address climate | The documents provide a detailed action plan for the local government's response to climate change, which includes specific actions, roles, timeframes, |

|  | change, they do not provide specific sources of evidence or data to justify these actions. For example, the documents mention the development of a 'whole-of-organisation' approach to climate change, the implementation of a Climate Emergency Strategy, and the embedding of climate change considerations into Council plans and policies. However, they do not provide specific evidence or data to support these actions or to demonstrate how they will effectively address climate change. Furthermore, while the documents mention the use of an annual monitoring and reporting program, they do not provide details on the specific metrics or indicators that will be used to measure progress or effectiveness. Therefore, while the documents outline a comprehensive approach to addressing climate change, they do not provide well-sourced evidence to justify their climate targets and actions. | responsibilities, outcomes, and indicators. The plan is comprehensive and covers a wide range of areas, including leadership and governance, community wellbeing and culture, business and economy, energy and buildings, biodiversity and open space, transport and movement, assets, infrastructure and land use planning, and waste and resources. The plan also includes a commitment to ongoing monitoring, reporting, and review of its implementation, with an annual progress summary report to be made public on the Council's website. The plan also includes a commitment to embed consideration of climate change into the development of the Council's 2021-25 Council Plan and supporting strategic documents. This suggests a well-sourced and justified approach to its climate targets and actions. |
| --- | --- | --- |
| 0.54 (median) | The documents do not explicitly discuss how to equitably share the benefits and opportunities of a safe climate. While the policy does mention the importance of considering 'equity implications of the adaptation option for all potential affected stakeholders', it does not provide specific details on how this will be achieved. The policy also outlines various actions to be taken in response to climate change, such as raising awareness among stakeholders, implementing adaptation solutions, and monitoring high impact invasive species. However, these actions are not directly linked to the equitable sharing of benefits and opportunities. The policy also does not provide a clear definition of what it considers to be 'equitable outcomes', nor does it outline specific strategies for achieving these outcomes. Therefore, based on the provided | The document does discuss the equitable sharing of benefits and opportunities in the context of climate change, although it does not explicitly use the phrase 'equitable sharing of benefits and opportunities of a safe climate'. The policy outlines the importance of considering the 'equity implications of the adaptation option for all potential affected stakeholders' when developing adaptation actions. This suggests a commitment to ensuring that the benefits and opportunities arising from these actions are distributed fairly among all those affected. Furthermore, the policy includes actions aimed at improving community resilience and social connectedness, and promoting education programs for business |

| | documents, it cannot be conclusively stated that the policy explicitly discusses how to equitably share the benefits and opportunities of a safe climate. | and household resilience. These actions indicate an intention to ensure that all members of the community have the opportunity to benefit from the policy's climate change adaptation measures. |
|---|---|---|

## Patterns in paired answers with opposite findings

Qualitative analysis of the set of paired answers with opposite findings highlights some patterns which illustrate aspects of GPT-4's operation.

### *Evidence vs interpretation*

The example in Table 28 illustrates a pattern that can be seen throughout the large-scale analysis (also noted in section 3.5, 'Variability'). When PALLM's finding is false, its long answer is more likely to refer to 'specific details', 'evidence' and 'explicit' discussion, conveying that the finding was false because of a lack of rigorous evidence. When PALLM's finding is true, it more often uses words and phrases such as 'suggests' or 'indicates', and appears more likely to 'read between the lines' of the document and make interpretations that lead to a true finding.

Table 29 shows the log-likelihood ratio (described in section 3.5) for words which illustrate this pattern. Across the two corpora of PALLM's 'true-finding' and 'false-finding' long answers, 75% of words have a log-likelihood ratio of less than 6.0, thus the high values for the words in Table 29 indicate that these words are strongly associated with one corpus over the other.

*Table 29: Selected words related to evidence or interpretation, and their log-likelihood ratio in long answers linked to true or false findings*

| More likely to appear in answers with 'true' finding | More likely to appear in answers with 'false' finding |
|---|---|
| indicates: 159.35 | explicitly: 1001.576 |
| suggests: 140.583 | evidence: 379.897 |
| indicating: 109.596 | specifically: 236.672 |
|  | explicit: 165.651 |
|  | specific: 18.923 |

*Past vs future*

Question 14 asks if the document contains "plans to conduct research in the local community". During the validation process, one evaluator noted that PALLM's response referred to the community engagement processes that were conducted during development of the plan, which strictly speaking should not be considered as future plans to conduct further research (I-B-1). The same misinterpretation is evident in one PALLM response with a positive finding for this question: "The documents do describe plans to conduct research in the local community to inform climate actions. The local government has undertaken a community survey to understand the concerns and ideas of the community regarding climate change" (PALLM, Buloke, 2024-03-13b). But this is specifically rejected in another response with a negative finding: "The documents do mention a community survey conducted by Ndevr Environmental, but this appears to be a past activity used to inform the drafting of the plan, rather than a future plan for research" (PALLM, Buloke, 2024-03-13a).

*Judgement*

The analytical process requires a degree of judgement to be applied, and just as human analysts may interpret the same text differently, PALLM can infer different conclusions from a document at different times. In its response to Question 6, for a document which states that consideration of climate risks "must be embedded across Council's services, strategies, policies and processes" (City of Greater Bendigo, 2021, p9), PALLM justified a true finding by stating that the above sentence "indicates a clear directive for all council activities to be aligned with the climate policy" (PALLM, Greater Bendigo, 2024-04-13b). However, in a negative finding to the same question, it noted "this [statement] is not the same as requiring all activities to align with a specific climate policy" (PALLM, Greater Bendigo, 2024-04-13a).

The wording of Question 12 - "Does the document provide well-sourced evidence to justify its climate targets and actions?" - seems particularly vulnerable to inconsistent interpretation, leading to variable output. This question was the target of six of the 22 responses with opposite findings. As can be seen in row one of Table 28, PALLM justifies a positive finding by listing the various characteristics of the policy, and its comprehensive nature, implying that the content of the policy itself constitutes 'well-sourced evidence'. The long answer for the negative finding focuses on the lack of externally-sourced evidence or data which would support the policy's proposed actions.

As well as the six findings for question 12, another six of the 22 opposite-finding responses were for questions 13-15 in the 'Planning for Informed Action' attribute, which relate to the evidence base and evaluation of the policy. Questions linked to this attribute formed over half of the inconsistent responses found in two executions. It seems that questions related to this subject involve more judgement and thus are more open to an inconsistent finding.

This section has discussed the small number of inconsistent findings in some depth, because they are a useful way to illustrate certain aspects of PALLM's operation. Inconsistent findings form only 1.5% of PALLM's responses, and in almost all cases PALLM has been demonstrated to respond consistently over time. The 22 inconsistent findings were excluded from the dataset used in the large-scale analysis presented in section 4.3.

## 4.5 Approaches to policy alignment

While the focus of this study is the large-scale analysis enabled by computational techniques, it is also possible to use NLP-based tools to provide insight into deeper aspects of policy, by using generated analysis "as a starting point for further, detailed examination" (Sachdeva et al, 2022). The binary finding returned by PALLM provides a simple answer to a complex question, where a qualitative assessment would provide much richer insight. PALLM's long answers provide summaries, quotations and examples which assist the researcher in conducting such an assessment.

As an example of the more nuanced analysis enabled by PALLM, this section presents a thematic analysis of 'Prioritisation of action', attribute three of the CEM framework. The questions for this attribute asked firstly whether a policy document shows evidence of prioritising a climate emergency response above other policies; and secondly, whether the document states that all council activities must be aligned with climate policy. For most councils, PALLM gave negative findings for these questions (answering positively for 24.4% of CED councils, and only 7.9% of non-CED). However, its long answers provide many examples of policy actions which relate to prioritisation and policy alignment, and this analysis describes some common themes. Such an analysis could be conducted for every attribute in the CEM framework, but it is only within the scope of this study to examine one.

Almost all climate policies or environmental sustainability strategies acknowledge that considerations of climate change must be incorporated into decisions and processes across all council activities. The Local Government Act 2020 requires councils to promote the sustainability of their district, including consideration of climate mitigation and addressing climate risks (DELWP, 2020), so it is necessary for policies to address this requirement. The approach to policy alignment is expressed in different ways for different councils, but a few themes can be distinguished.

*Integration into decision-making*

An estimated 75% of councils describe how their climate policy relates to other policies and plans the council has produced, and how it relates to high-level strategies. Commonly a list of related policies and strategies will be provided, and in some cases a diagram is used to indicate their hierarchy, usually with an overarching Council Plan at the top. There is frequently also an acknowledgement that climate change must be considered in all council planning and decision-making. For example, the Campaspe Shire Council's "Environment Strategy", which is not a solely climate-focused policy and lists climate as one of four 'themes', states that the council must "[i]ncorporate consideration of climate change and relevant state and national plans into strategic planning. The long-term adverse consequences of climate change for future generations are incorporated into Council planning, decisions and actions" (Campaspe Shire Council, 2022, p19).

*Embedding resilience*

More than half the councils in the dataset (39 in total, 28 CED and 11 non-CED) refer to 'embedding' climate considerations in council activities. These policies describe in more detail how climate should be integrated into operations, but for some the focus is primarily on increasing climate resilience across all council services and activities. Essentially this refers to risk management, incorporating risk assessment and adaptation principles into decision-making. For example, Buloke Shire Council's "Climate Action Plan" states a goal to "Integrate climate into Council operations" which includes specific actions including "Include climate scenario and risk profile in Councillor briefing packs and staff induction" and "Use the How Well Are We Adapting tool to monitor impacts of climate change on Council services and develop responses" (Buloke Shire Council, 2021, p4). The focus is on ensuring that all council planning and decision-making anticipates potential future impacts of climate change.

*Climate lens*

The language used to describe 'embedding' climate into general operations often refers to a "climate lens" (or variations on that phrase), or "climate thinking". Seventeen of the 95 documents (ten from CED councils, four from non-CED) use some form of this phrase. For example, the Brimbank City Council's "Climate Emergency Plan" describes six principles of their "Climate Emergency Lens" which include aspects of decision-making, risk management, equity, engagement and compatibility with other governments (Brimbank City Council, 2020, p8). Some policies note that applying a climate lens requires additional investment in staff capacity: "While the scientific evidence base for climate change and emission reductions is extensive and clear, applying the climate lens requires a change in individual mindsets" (Hume City, 2023, p23).

The Indigo Shire "Climate Emergency Strategic Action Plan" describes its approach to embedding the climate emergency into its operations as "a red-carpet, not a red-tape approach … rolling-out the welcome mat for ideas and actions" and expects consideration of the climate emergency to affect everything from the Council Plan "right through to how individual staff members interact with the public" (Indigo Shire, 2020, p2).

*Embedding action*

A small number of documents describe a "whole of council" approach to climate response which embeds consideration of not only climate risk but climate action across council operations. With this approach, policies consider not only the impact of climate change on council activities, but also the impact of council activities on the climate.

Some policies encourage application of a climate lens to procurement and investment decisions, regarding not only risk but also mitigation opportunities. The Indigo Shire plan commits to "including consideration of environment/climate change criteria when evaluating quotations for applicable goods and services" (Indigo Shire, 2020, p3). The Bayside City Council's "Climate Emergency Action Plan" encourages "the future cost of inaction" to be factored into resourcing decisions, and suggests that "[n]on-essential functions and consumption may be curtailed or rationed through an emergency lens prioritisation process" (Bayside City Council, 2020, p7).

Adjustments to standard documents and processes are described as part of a climate lens approach, such as adding a Climate Impact Assessment section to planning and reporting templates, including climate change in staff induction programs and position descriptions, and publicly reporting on emissions in regular reporting cycles.

In some councils, the climate policy is given a key strategic focus: for example, the Brimbank plan "sits at the highest level of Council's strategic planning framework, above and informing the four-year Council plans that are established at each new Council term" (Brimbank City Council, 2020, p6). Integration of climate policy into the council's strategic plan is the third of five recommended steps for local governments to enter 'emergency mode' (Spratt, 2019). At least four councils aim to include climate commitments in KPIs for the CEO and executive team, for example "Embed delivery of the Climate Action Plan within the CEO's contract / performance plan" (Nillumbik Shire, 2022, p22).

All actions cited as examples of "embedding action" are from councils who have passed a CED, and offer signs that the pressure to deliver a strong climate emergency response may be leading to an increasing strategic and operational focus on climate as a priority for local governments.

## 4.6 Summary

This chapter has presented the results which address my three research objectives. Section 4.2 discussed the outcomes of an evaluation process for PALLM, which found that policymakers generally showed high agreement rates with its responses, while section 4.4 explored some patterns which illustrate aspects of LLM behaviour (RO1, RO2). Section 4.3 presented the results of a large-scale analysis exploring the influence of the climate emergency movement on climate policy in Victoria, finding that CED councils show more prevalence of attributes relating to urgency, prioritisation, and equity and social justice; and section 4.5 offers a qualitative assessment of one aspect of policy (RO3). These results illuminate aspects of LLM capabilities and of the impact of the climate emergency movement, and these will be discussed in the next chapter.

# 5. Discussion

## 5.1 Introduction

In this project I have set out to answer research questions about the capabilities of LLMs as a tool for policy analysis, and about the influence of the Climate Emergency movement on climate policy in local governments in Australia. The research questions have been investigated through a technical solution named PALLM, using GPT-4 to answer questions on climate policy documents. This project has generated insights both methodologically, in the use of GPT-4 for policy analysis, and substantively, in demonstrating differences between climate policies from councils which passed a climate emergency declaration and councils which did not.

## 5.2 RQ1: What are the current capabilities of large language models in assessing policy documents?

This research finds that LLMs are capable of high-level policy analysis, performing particularly well at summarising and selecting examples of policy actions, but are limited by the ability to process complex or context-dependent information, and by a lack of reliable attribution.

Research on LLMs is fast-evolving and rapidly moving into new areas, but to date has not demonstrated a robust and well-evaluated solution for policy analysis. In climate research, the focus on LLMs has primarily been on QA systems which can interact with scientific and factual data, although other computational tools such as topic modelling have been used to analyse policy documents at scale.

This work has examined the potential of a QA/RAG system using GPT-4 to answer complex questions about climate policy documents, and has evaluated the output of this system with help from policymakers. This has led to a number of insights regarding the strengths and limitations of the PALLM system. The strengths are summarised here as relating to scale and parametric knowledge, while the limitations are categorised into four groups related to context, complexity, variability and attribution.

## Strengths

### *Analysis at scale*

Using PALLM, it is possible to apply a conceptual framework to 95 policy documents in two days, far exceeding what an individual researcher could do manually in this time. PALLM's output provides quantitative and qualitative data which the researcher can work with in many ways.

In this study, quantitative data was generated by evaluative questions with a structured response format. While a binary answer may seem too reductive for some complex questions, it enables straightforward data analysis, and the validation process generally found high agreement rates with PALLM's findings.

PALLM also produces qualitative data by asking GPT-4 to generate an explanation for its finding. This will typically include relevant examples of policy actions, and summaries of high-level themes of the document. PALLM proved effective at identifying and restating important aspects of the document, and selecting relevant information to include. These 'long answers' provide a useful summary and pointer to further information, which the researcher can use for further qualitative analysis (Sachdeva et al, 2022).

PALLM's level of accuracy in extracting and interpreting information from documents was found by evaluators to be high. Although some misinterpretations occurred due to lack of context, this study did not find any evidence of hallucinations in GPT-4's long answers, such as incorrect information, or information that was not present in the source documents. This does not guarantee that no hallucinations are present, as the large-scale analysis results have not been systematically evaluated, but none were detected during validation or manual analysis.

*Parametric knowledge*

For simple RAG systems, the LLM's task is primarily one of reading comprehension and information extraction, where a retrieval system identifies relevant excerpts and the LLM extracts and summarises the important data points. However, for more complex questions in policy analysis, the LLM component of PALLM must be able to understand the context of the selected document excerpts and be familiar with the typical content of such documents. The high agreement rates found in the validation process suggest that GPT-4 performed effectively in this regard and indicate that the parametric knowledge derived from its training is sufficient for climate policy analysis tasks.

Interview and survey responses from evaluators indicated that the long answers generated by GPT-4 demonstrated a good understanding of basic climate science and policy, and usually correctly identified mitigation and adaptation actions. If no such actions were found, GPT-4 would sometimes suggest typical actions that could have been included, which evaluators found credible. It successfully classified mitigation actions into the suggested categories of 'technological solutions' and 'behaviour change', and could explain the mechanism by which specific policy actions would help to reduce emissions or increase resilience when the document did not provide such detail.

## Limitations

*Context*

Given the fixed limit of 8192 tokens on each interaction with the GPT-4 model, only selected excerpts of the document could be provided in each prompt, meaning the model did not have access to the full text. Thus, when PALLM's findings were incorrect or the long answer omitted important information, there were two possible reasons: (1) the retrieval system did not select this information for inclusion in the prompt, and/or (2) the information was included but GPT-4 did not recognise it as important. Additionally, the text extraction process could not identify data encoded in images, and presented data formatted in tables separately from relevant column headings. This meant that GPT-4 did not necessarily have the most appropriate information with which to answer a question.

The lack of context for document excerpts occasionally meant that GPT-4 misinterpreted the origin of statements – for example, PALLM's finding on question 5 for the Greater Geelong "Climate Change Response Plan" (2021) stated in part that the policy prioritises climate action above other policies, "even suggesting that climate considerations should take precedence over financial outcomes" (PALLM, Greater Geelong, 2024-04-13b). This statement seems to be based on the sentence "Prioritising climate over financial outcomes in all decision making was also highlighted" (Greater Geelong City Council, 2021, p19) which is a summary of community feedback on the draft plan, not an explicit policy intention.

The use of a retrieval system also limited PALLM's ability to accurately answer questions relating to the overall structure or purpose of a policy document, illustrating what Thulke et al describe as "an inherent trade-off between factuality and abstractiveness" in RAG systems (2024, p17). This is an area of rapid development and advances in document parsing (such as Lin, 2024) are likely to greatly improve the performance of retrieval systems for this purpose.

*Complexity*

The questions in PALLM's question set can be considered "complex questions" as defined by Daull et al (2023). Complex QA requires a multi-step resolution process, and draws on cognitive skills such as analysis and inference, including the ability to cope with ambiguity and nuance (Daull et al, 2023).

The level of confidence rating assigned to each question in section 4.3 reflects PALLM's effectiveness at complex QA. While the rating is an imprecise measure, it provides insights into the types of questions which PALLM is more successful at answering. Questions with Tier 1 confidence included those which focus directly on the language of the document, such as "Does the document explicitly call for rapid and urgent action on climate change?" and "Does the document explicitly discuss the impact of climate change on vulnerable communities?". Answering these questions requires detecting references to concepts of 'urgency' and 'vulnerable communities', which are relatively clear and unambiguous. Other questions with high confidence were those with similar findings across the dataset, i.e. where most councils receive the same finding (for example, questions on the presence of mitigation or adaptation actions).

Questions with Tier 3 confidence included those which require multiple steps to execute and judgement on nuanced concepts, such as "Does the document provide well-sourced evidence to justify its climate targets and actions?". Answering this question requires the execution of several logical steps (What are the targets and actions? What evidence is provided to justify them? What sources are cited for this evidence?) and evaluation against unspecified criteria (How do we know if evidence is 'well-sourced'?). The questions in the 'Planning for informed action' attribute were particularly prone to low levels of confidence, caused by low agreement rates in validation and more inconsistent findings across executions. This attribute examines aspects of policy development and implementation, evidence for which must be inferred rather than detected solely by the text in the document. Other questions in this attribute with low confidence include "Does the plan include specific measurable criteria to evaluate the success of its proposed actions?" and "Does the document show evidence of innovation and policy experimentation in climate action?", both of which require contextual understanding and judgement on what counts as 'measurable' or as 'innovative'.

*Variability*

As discussed in chapters 3 and 4, some variability is unavoidable due to the inherently non-deterministic nature of LLM operations. More complex questions are more prone to inconsistent findings, and fine-tuning or customisation of the GPT-4 model used in PALLM could assist in reducing these.

As discussed in section 4.4, PALLM tends to rely on 'evidence' to justify negative findings, and 'interpretation' for positive findings. The acceptable degree of interpretation is hard to define, because a motivated reader can detect presence of an attribute by 'reading between the lines' and making assumptions. During initial development, GPT3.5 was found to be very prone to this behaviour, and seemed highly motivated to answer questions with a positive finding (and accompanying justifications which sometimes stretched credibility). Once PALLM moved to GPT-4, this behaviour improved significantly, and the changes to the prompt detailed in section 3.5 also made GPT-4 a less 'generous' interpreter, more reliant on explicit evidence.

Answers to complex questions are themselves complex, and some level of inference will usually be required. Human readers may disagree with GPT-4's findings, as any analyst may disagree with another. There is often no objectively correct answer to these questions, which is why it is critical for analysis to include detailed justification and evidence for its assertions.

*Attribution*

The attempt to provide reliable attribution for PALLM's output was the least successful aspect of this project. The prompt for each question asked GPT-4 to choose a representative quotation from the document to accompany each positive finding, and PALLM then applied a verification step to ensure that the quotation was genuine. Problems with this process included the following:

- The quotations selected by GPT-4 were sometimes very appropriate to the topic, but at other times their relevance was unclear.

- The quotations returned were often not verbatim from the document, but consisted of unrelated sentences concatenated together, or partial sentences with an invented ending.

- The verification process applied by PALLM could filter out some invalid quotations, but not all.

Additionally, the long answers generated by GPT-4 sometimes contained long phrases or sentences which were quoted verbatim from the document without attribution.

Technical improvement within PALLM could improve its verification process, but the remaining issues stem from the inherent tendency of LLMs to hallucinate (Zhao et al, 2023), and are more challenging to solve.

Reliable attribution would significantly contribute to the usefulness of PALLM as a policy analysis tool, as it allows quick verification of PALLM's findings and increases its credibility through greater reliance on the source document (Bohnet et al, 2022). Recent work by Susnjak et al (2024), which proposes a token-based system of 'knowledge markers' to enable auditing and verification of source information in LLM-generated text, could be valuable in this regard.

## 5.3 RQ2: To what extent has the language and priorities of the Climate Emergency movement influenced Australian local government policy?

This study finds that the language and priorities of the Climate Emergency movement are visible within local government climate policy in Victoria, particularly regarding the sense of urgency, prioritisation of climate, and attention to equity and social justice displayed within policy documents.

The aims and impact of the Climate Emergency movement have been described and critiqued by several researchers (Chou, 2020; Davidson et al, 2021; Howarth et al, 2021; Greenfield, 2022; Salvia et al, 2023). At the movement's peak in 2019-2020, its primary goal was to push for accelerated action in climate governance, moving from 'business as usual' into 'emergency mode' (Salamon, 2019; Spratt, 2019).

The dataset assembled for this project includes a number of climate policies and plans that have been produced by councils after their declaration of climate emergency. Compared to non-CED, CED councils have more recent policies and are more likely to have a stand-alone climate policy. While Salvia et al (2023) found that for two-thirds of the cities in their study, development of a local climate plan preceded a declaration of climate emergency, the findings of this study indicates that councils which passed a CED were more likely to subsequently produce a stand-alone climate plan. This concurs with Harvey-Scholes (2019) who found that 93% of UK CEDs contained a commitment to produce a Climate Action Plan. Regardless of the chronology, this study's findings concur with Salvia et al (2023) that the attention brought to climate change by the Climate Emergency and School Strike for Climate movements has motivated local governments to respond with declarations and climate action plans.

The large-scale analysis generated by PALLM provides a comprehensive view of climate policy across local governments in Victoria. While other studies have used computational tools to analyse large numbers of policy documents (Sachdeva et al, 2022; Salvia et al, 2023; Hsu & Rauber, 2021), these have tended to focus on specific aspects such as net zero targets or particular types of mitigation action. Studies using a robust conceptual framework have generally focused on a small number of case studies (Greenfield et al, 2022; Howarth et al, 2021). This study's contribution is to combine these approaches by applying a framework that examines multiple aspects of climate policy, at a broad enough scale to outline the policy landscape.

The following sub-sections will discuss the study's findings regarding the attributes of the CEM framework, including those with generally high and low prevalence, and those with significant relative difference between CED and non-CED cohorts.

## High-prevalence attributes

The results for each attribute of the CEM framework, as presented in section 4.3, show that CED councils score more highly on every attribute than non-CED councils. In some attributes both cohorts score highly, particularly in the attributes of 'Purpose of action', 'Social mobilisation', 'Restoring a safe climate' and 'Adapting to a changing climate'. A high score for 'Purpose of action' is unsurprising given that the dataset was limited to policies either entirely climate-focused or with significant climate content. Beyond that, the results indicate that most local government climate policies in Victoria include discussion of community education and engagement, mitigation actions and adaptation actions. This confirms and extends the work of Salvia et al (2023) and Harvey-Scholes (2019), who surveyed zero carbon targets in CEDs in the UK and Italy, although the presence of specific goals or targets is not quantified in this study.

The finding of high prevalence for 'Social mobilisation' in both cohorts contrasts with Davidson et al (2024), which did not find strong evidence for this attribute. The wording of the question used for this attribute may have resulted in a broader conception of "social mobilisation", which could be seen as present whenever the policy refers to any form of community education or engagement.

### Low-prevalence attributes

The attributes with the lowest mean prevalence in both cohorts were 'Prioritisation of action' (18.1%) and 'Equity and social justice' (36.6%), both of which reflect concerns of the broader climate movement, but which are not commonly expressed in policies narrowly focused on direct climate impacts and emissions reduction. This low prevalence of these attributes correlates with findings from previous studies in the Victorian context by Davidson et al (2020) and Davidson et al (2024).

### Attributes with high relative CED/non-CED difference

Together with 'Urgency of action', the attributes of 'Prioritisation' and 'Equity and social justice' showed the greatest relative difference between CED and non-CED councils, and are where the language and priorities of the Climate Emergency movement can be seen most clearly. In all three of these attributes, the mean score among CED councils was over 30% higher than among non-CED councils. The themes of urgency, prioritisation and social justice are often evident in CED statements, as identified by Greenfield et al (2022) who noted themes of urgency, vulnerability, leadership and engagement in climate emergency declarations. This study builds on that finding by detecting these attributes in policies as well as declarations.

*Urgency of action*

'Urgency of action' is the attribute where 'emergency' discourse could be expected to be most evident, and this attribute shows the largest absolute difference between cohorts, present in 84.1% of CED council policies but only 56.2% of non-CED. Greenfield et al (2022) describes the theme of urgency as relating to both 'immediacy' (including reference to timeframes) and 'action' (to move beyond symbolic statements), and both of these aspects are evident in PALLM's long answers related to this attribute. Davidson et al (2024) also find evidence of urgency in topics related to $CO_2$ drawdown and moving away from business as usual.

*Equity and social justice*

The attribute of 'Equity and social justice' is more prevalent within CED than non-CED councils, though it was absent in more than half of policies across both cohorts. This agrees with Davidson et al (2024) which found limited evidence for this attribute in examined policies. Greenfield et al (2022) noted that the theme of 'vulnerability' in CEDs was expressed in two ways: the locale's geographic vulnerability to climate impacts, and the vulnerability of particular communities who would be disproportionately affected by climate change. PALLM's long answers in responses for this attribute frequently referred to the second aspect but did not mention the first, indicating that PALLM was able to distinguish between these two types of vulnerability.

*Prioritisation of action*

Prioritisation of climate action is a key demand of the climate emergency movement (Spratt, 2019). The quantitative analysis within this study found a low prevalence for prioritisation of climate, but the high-level questions posed by PALLM do not necessarily detect the subtle ways in which climate considerations can be embedded across an organisation. A deeper qualitative assessment of climate prioritisation and alignment in the dataset, described in section 4.5, found that CED council policies give significantly more attention to strategic focus on climate, and suggest actions to embed climate action and climate risk assessment across the organisation, while non-CED policies show little mention of prioritisation. The association of CEDs with greater discussion of strategy and governance lends support to a previous finding that the presence of language related to governance in policies predicts more ambitious net-zero targets (Sachdeva et al, 2022).

Studies using the CEM framework (Davidson et al, 2020; Greenfield et al, 2022) have found the attribute of prioritisation to have the least evidence supporting its presence. This work shows that the process of embedding climate considerations into all council activities is underway at CED councils, but does not refute Greenfield's conclusion that "[i]f prioritization of climate change action over other local government action is taken as the threshold for being in emergency mode, then the declarations of CEs studied here are not reaching this threshold" (2022, p10).

These findings help to answer the question raised by Greenfield et al (2022), Chou (2020), and Howarth et al (2021), among others, as to whether CEDs are a purely symbolic action or have real-world impact. While the higher prevalence of CEM attributes among CED councils indicates correlation and is not proof of causality, it does suggest that climate emergency declarations are associated with measurable differences in policy, and that they operate beyond the realm of symbolic politics (Greenfield et al, 2022).

## 5.4 Summary

The current capabilities of LLMs for policy assessment include analysis of a range of attributes, summarising policy data, selecting relevant examples and answering complex evaluative questions. There are limitations to these capabilities, particularly regarding complexity and attribution, but their general accuracy has been validated by policymakers.

The extent of the impact of the Climate Emergency movement on policy cannot be precisely quantified, but it is clear that councils which have passed a CED are more likely to produce a stand-alone climate plan, and their climate policies show more attention than non-CED councils to urgency, to prioritisation of climate considerations, and to social justice issues.

# 6. Conclusion

In this study, I have completed the three stated research objectives of building and validating a technical solution to analyse policy documents with GPT-4, assessing its capabilities, and conducting an analysis of local government climate policies. This has enabled me to answer the two research questions, generating some methodological and substantive insights, and to suggest some approaches for future research.

RQ1. The use of PALLM in this project enabled me to conduct a large-scale analysis of 95 policy documents. The validation process found that policymakers generally had high levels of agreement with PALLM's findings and reported that it could be a useful tool for their work. I conclude that current LLMs have significant capability to enable both quantitative and qualitative research on large datasets of documents.

RQ2. Using a combination of quantitative data analysis and qualitative assessment of source documents, aided by PALLM's synthesis of specified topics, this study has demonstrated a clear difference between the policies produced by CED and non-CED councils, most notably in the attributes of urgency, prioritisation, and equity and social justice. While this should not be taken to infer a causal relationship between CEDs and characteristics of subsequent policy documents, the language and priorities of the Climate Emergency movement are visible in both types of document in CED councils.

There are of course limitations to what the study of documents can achieve. Analysing the text of policy documents does not shed light on how those policies were developed, how they are implemented, or what their outcomes are. As one evaluator noted, while an AI-generated analysis can describe the intention behind a policy document, "it could be usefully enhanced by having a conversation [about] what's different in the journey that you've had, because I think that's where the learnings are for a lot of organisations" (I-S-2).

The scores assigned to councils by PALLM based on their policy documents are necessarily general and imprecise, and do not take into account the numerous other actions and characteristics of a local government organisation. The scores are not intended to rank councils against each other and small differences between the overall score, or between different attributes, should be disregarded. The importance of the analysis is in its ability to provide a snapshot of the broad policy landscape.

It is also important to reiterate that PALLM's computational nature does not make its output deterministic or 'objective'. While some evaluators perceived 'objectivity' as a key strength of PALLM (see section 4.2), evidence of variability in PALLM's findings demonstrates that it is not uncovering an objective truth, but rather constructing an argument based on probabilistic choices.

With those caveats in mind, the use of LLM-based analysis opens up an exciting new space in policy research, and could enable individual researchers to pursue broader and deeper analytical goals. There are many directions in which future research could expand on this study.

## Future research recommendations

The technical ecosystem around LLMs is developing rapidly, and GPT-4 and associated tooling like Langchain have more capabilities today than at the beginning of this study. The newly-released OpenAI model GPT-4o has the ability to extract data from images within a PDF document, as well from other media types. Other OpenAI models such as GPT-4 Turbo (or other types of LLM) have a much larger context buffer and could process most policy documents without the use of a retrieval system, which would avoid many of the context issues discussed in Chapter 5. Fine-tuning and customisation of GPT-4 could enhance the consistency of PALLM's answers by clarifying the criteria which it uses to make judgements about complex concepts. Deeper investigation in this area could also illuminate important aspects of LLM operation.

Further technical development within PALLM could also assist with improving attribution verification, and additional revision of prompt instructions and question wording may improve its output.

With or without additional development, PALLM could be used to conduct further analysis on policy documents. The scope of the policy dataset could be increased as far as time and financial resources permit, to a national or international scope. The questions associated with the CEM framework could be substituted or enhanced, and additional questions could explore other aspects of policy. Although this study has not explored the use of LLMs in systematically extracting information (such as net zero targets, timeframes or particular mitigation actions) from unstructured text, this has been demonstrated in other work (Vaghefi et al, 2023; Kraus et al, 2023) and the results could be integrated into PALLM's output.

As demonstrated in section 4.5, PALLM also acts as research assistant, enabling qualitative assessment of unstructured text by selecting and summarising relevant information. Within the present dataset, this could be used for a deeper exploration of certain attributes of climate policy, such as the different ways in which social mobilisation is characterised, or examples of policy innovation. PALLM could examine policy documents to detect the presence of common themes identified in CED documents (Greenfield et al, 2022), or attempt to verify the correlation of themes identified by Sachdeva et al (2022) with stronger net zero targets.

While the momentum of the Climate Emergency movement has slowed since its peak in 2019, its impact on policy and discourse remains visible, even as the urgency of the climate crisis continues to grow. With a rapidly-expanding computational toolkit, researchers will be better equipped to track the fast-moving policy landscape.

# Appendix: Victorian local governments and climate emergency declarations

| Local Government Area | Date of Climate Emergency Declaration | Number of policy documents in dataset |
|---|---|---|
| Alpine Shire Council | 9 November 2021 | 2 |
| Ararat Rural City Council |  | 0 |
| Ballarat City Council | 21 November 2018 | 1 |
| Banyule City Council | 7 October 2019 | 1 |
| Bass Coast Shire Council | 21 August 2019 | 1 |
| Baw Baw Shire Council |  | 0 |
| Bayside City Council | 17 December 2019 | 1 |
| Benalla Rural City Council |  | 1 |
| Boroondara City Council | 27 September 2021 | 3 |
| Brimbank City Council | 25 June 2019 | 3 |
| Buloke Shire Council |  | 1 |
| Campaspe Shire Council |  | 1 |
| Cardinia Shire Council | 16 September 2019 | 3 |
| Casey City Council |  | 1 |
| Central Goldfields Shire Council |  | 1 |
| Colac-Otway Shire Council |  | 1 |
| Corangamite Shire Council |  | 1 |
| Darebin City Council | 5 December 2016 | 2 |
| East Gippsland Shire Council |  | 1 |
| Frankston City Council | 18 November 2019 | 1 |
| Gannawarra Shire Council |  | 1 |
| Glen Eira City Council | 5 May 2020 | 1 |
| Glenelg Shire Council |  | 1 |
| Golden Plains Shire Council | 27 July 2021 | 1 |
| Greater Bendigo City Council |  | 1 |

| Greater Dandenong City Council | 28 January 2020 | 2 |
| --- | --- | --- |
| Greater Geelong City Council | 25 February 2020 | 1 |
| Greater Shepparton City Council | 31 March 2020 | 1 |
| Hepburn Shire Council | 17 September 2019 | 2 |
| Hindmarsh Shire Council |  | 1 |
| Hobsons Bay City Council | 8 October 2019 | 1 |
| Horsham Rural City Council |  | 1 |
| Hume City Council |  | 1 |
| Indigo Shire Council | 30 July 2019 | 2 |
| Kingston City Council | 28 January 2020 | 3 |
| Knox City Council | 27 September 2021 | 1 |
| Latrobe City Council |  | 1 |
| Loddon Shire Council |  | 1 |
| Macedon Ranges Shire Council | 24 March 2021 | 1 |
| Manningham City Council | 28 January 2020 | 1 |
| Mansfield Shire Council |  | 1 |
| Maribyrnong City Council | 19 February 2019 | 1 |
| Maroondah City Council |  | 2 |
| Melbourne City Council | 16 July 2019 | 1 |
| Melton City Council |  | 2 |
| Merri-bek City Council | 12 September 2018 | 1 |
| Mildura Rural City Council | 26 February 2020 | 1 |
| Mitchell Shire Council | 20 September 2021 | 1 |
| Moira Shire Council |  | 1 |
| Monash City Council |  | 1 |
| Moonee Valley City Council | 8 October 2019 | 1 |
| Moorabool Shire Council |  | 1 |
| Mornington Peninsula Shire Council | 13 August 2019 | 2 |
| Mount Alexander Shire Council | 17 December 2019 | 1 |
| Moyne Shire Council | 22 October 2019 | 2 |

| Murrindindi Shire Council | | 1 |
|---|---|---|
| Nillumbik Shire Council | 26 April 2022 | 1 |
| Northern Grampians Shire Council | | 0 |
| Port Phillip City Council | 18 September 2019 | 1 |
| Pyrenees Shire Council | | 1 |
| Borough of Queenscliffe Council | 19 December 2019 | 1 |
| South Gippsland Shire Council | | 1 |
| Southern Grampians Shire Council | | 1 |
| Stonnington City Council | 17 February 2020 | 1 |
| Strathbogie Shire Council | 20 April 2021 | 1 |
| Surf Coast Shire Council | 27 August 2019 | 1 |
| Swan Hill Rural City Council | | 1 |
| Towong Shire Council | | 0 |
| Wangaratta Rural City Council | | 1 |
| Warrnambool City Council | 7 October 2019 | 1 |
| Wellington Shire Council | | 1 |
| West Wimmera Shire Council | | 0 |
| Whitehorse City Council | 12 September 2022 | 2 |
| Whittlesea City Council | | 1 |
| Wodonga City Council | | 1 |
| Wyndham City Council | | 2 |
| Yarra City Council | 7 February 2017 | 2 |
| Yarra Ranges Shire Council | 10 September 2019 | 3 |
| Yarriambiack Shire Council | | 0 |